\documentclass[journal,comsoc]{IEEEtran}
\usepackage[T1]{fontenc}

\ifCLASSINFOpdf
\else
\fi
\usepackage{cite}
\usepackage{amsmath,amssymb,amsfonts}
\usepackage{algorithm,algorithmic}
\usepackage{graphicx}
\usepackage{xcolor}
\usepackage{algorithmic}
\usepackage{graphicx}
\usepackage{caption}
\usepackage{subcaption}
\usepackage{amsmath}
\usepackage{csquotes}
\usepackage{tabu}
\usepackage{amsthm}
\usepackage{mathtools}
\DeclareMathOperator*{\argmin}{arg\,min}
\DeclareMathOperator*{\argmax}{arg\,max}

\newtheorem{definition}{Definition}

\newtheorem{thm}{Theorem}

\newtheorem{proposition}{Proposition}

\usepackage[cmintegrals]{newtxmath}
\begin{document}
\title{Risk-Aware Energy Scheduling for Edge Computing with Microgrid: A Multi-Agent Deep Reinforcement Learning Approach}

\author{Md.~Shirajum~Munir,~\IEEEmembership{Graduate~Student~Member,~IEEE,}
	Sarder~Fakhrul~Abedin,~\IEEEmembership{Member,~IEEE,}
	Nguyen~H.~Tran,~\IEEEmembership{Senior~Member,~IEEE,}
	Zhu~Han,~\IEEEmembership{Fellow,~IEEE,}
	Eui-Nam~Huh,~\IEEEmembership{Member,~IEEE,}
	and~Choong~Seon~Hong,~\IEEEmembership{Senior~Member,~IEEE}
\thanks{This work was supported by the Korea Institute of Energy Technology Evaluation and Planning(KETEP) and the Ministry of Trade, Industry \& Energy(MOTIE) of the Republic of Korea (No. 20199810100050).}
\thanks{Md. Shirajum Munir, Eui-Nam~Huh, and Choong Seon Hong are with the Department of Computer Science and Engineering, Kyung Hee University, Yongin-si 17104, Republic of Korea (e-mail: munir@khu.ac.kr; johnhuh@khu.ac.kr; cshong@khu.ac.kr).}
\thanks{Sarder Fakhrul Abedin is with the Department of Information Systems and	Technology, Mid Sweden University, 851 70 Sundsvall, Sweden, and also with the Department of Computer Science and Engineering, Kyung Hee University, Yongin 17104, South Korea (e-mail: sarder.abedin@miun.se).}
\thanks{Nguyen H. Tran is with the School of Computer Science,
The University of Sydney, Sydney, NSW 2006, Australia,(e-mail: nguyen.tran@sydney.edu.au).}
\thanks{Zhu Han  is with the Electrical and Computer Engineering Department, University of Houston, Houston, TX 77004 USA, and also with the Department of
	Computer Science and Engineering, Kyung Hee University, Yongin-si 17104, Republic of Korea,(e-mail: zhan2@uh.edu).}
\thanks{Corresponding author: Choong Seon Hong (e-mail: cshong@khu.ac.kr).}
\thanks{\textcopyright 2021 IEEE. Personal use of this material is permitted.  Permission from IEEE must be obtained for all other uses, in any current or future media, including reprinting/republishing this material for advertising or promotional purposes, creating new collective works, for resale or redistribution to servers or lists, or reuse of any copyrighted component of this work in other works.}}

\markboth{ACCEPTED ARTICLE BY IEEE Transactions on Network and Service Management, DOI: 10.1109/TNSM.2021.3049381}%
{Shell \MakeLowercase{\textit{et al.}}: Bare Demo of IEEEtran.cls for IEEE Communications Society Journals}
%



\maketitle

\begin{abstract} 
In recent years, multi-access edge computing (MEC) is a key enabler for handling the massive expansion of Internet of Things (IoT) applications and services. However, energy consumption of a MEC network depends on volatile tasks that induces risk for energy demand estimations. As an energy supplier, a microgrid can facilitate seamless energy supply. However, the risk associated with energy supply is also increased due to unpredictable energy generation from renewable and non-renewable sources. Especially, the risk of energy shortfall is involved with uncertainties in both energy consumption and generation. In this paper, we study a risk-aware energy scheduling problem for a microgrid-powered MEC network. 
First, we formulate an optimization problem considering the conditional value-at-risk (CVaR) measurement for both energy consumption and generation, where the objective is to minimize the expected residual of scheduled energy for the MEC networks and we show this problem is an NP-hard problem.
Second, we analyze our formulated problem using a multi-agent stochastic game that ensures the joint policy Nash equilibrium, and show the convergence of the proposed model. Third, we derive the solution by applying a multi-agent deep reinforcement learning (MADRL)-based asynchronous advantage actor-critic (A3C) algorithm with shared neural networks. This method mitigates the curse of dimensionality of the state space and chooses the best policy among the agents for the proposed problem. Finally, the experimental results establish a significant performance gain by considering CVaR for high accuracy energy scheduling of the proposed model than both the single and random agent models. 

\end{abstract}


\IEEEpeerreviewmaketitle

\section{Introduction}
\IEEEPARstart{M}{ulti-access} edge computing (MEC) enables a large amount of computational tasks for massive IoT-applications and background services to be executed by smart services \cite{IEEEhowto:Porambage_Survey_MA_IOT}, which requires more energy consumption as compared with normal execution of the wireless networks \cite{IEEEhowto:Mao_A_Survey_on_Mobile_Edge}. To handle the continual growth of energy consumption for wireless networks \cite{IEEEhowto:Abedin_Green_IoT}, renewable energy usage is essential for interrupt-free wireless network operation by reducing the dependency on non-renewable energy usage. Furthermore, research shows that a proper combination of energy generation (i.e., renewable, non-renewable, and storage) and distribution can save a significant amount of energy usage for radio access networks \cite{IEEEhowto:Han_grid_sdn}. Specifically, a jointly optimized demand-side management mechanism is claimed to save up to $18\%$ of the total energy usage in wireless networks \cite{IEEEhowto:Huang_Jointly_Optimizing_BS}. Therefore, renewable energy aware task scheduling and resource allocation for wireless networks infrastructure are necessary when considering energy consumption \cite{IEEEhowto:Wei_RL_Scheduling_Resource}. However, risk-aware energy scheduling is overlooked in a microgrid-powered MEC networks, where energy consumption of the MEC network strongly depends on the nature of MEC task fulfillment over time \cite{IEEEhowto:Munir_Edge_Microgrid}.  

In case of the physical deployment of MEC, the technical and business point of view are considered \cite{IEEEhowto:ETSI_White_deployment}. 
MEC facilitates various applications such as those that are related to smart cities, health care, smart agriculture, automotive, virtual reality (VR), and augmented reality (AR) \cite{IEEEhowto:Porambage_Survey_MA_IOT}, along with user specific requirements.
Consequently, MEC is already included as an essential component in various smart infrastructures such as smart cites and smart factories. Meanwhile, microgrids have been considered prominent in those MEC infrastructures \cite{IEEEhowto:Microgrid_Hernandez} \cite{IEEEhowto:Microgrid_Calvillo}. Thus, microgrid can be an effective energy supplement to MEC. 

\begin{figure}[!t]
	\centering 
	\includegraphics[width=8.9cm]{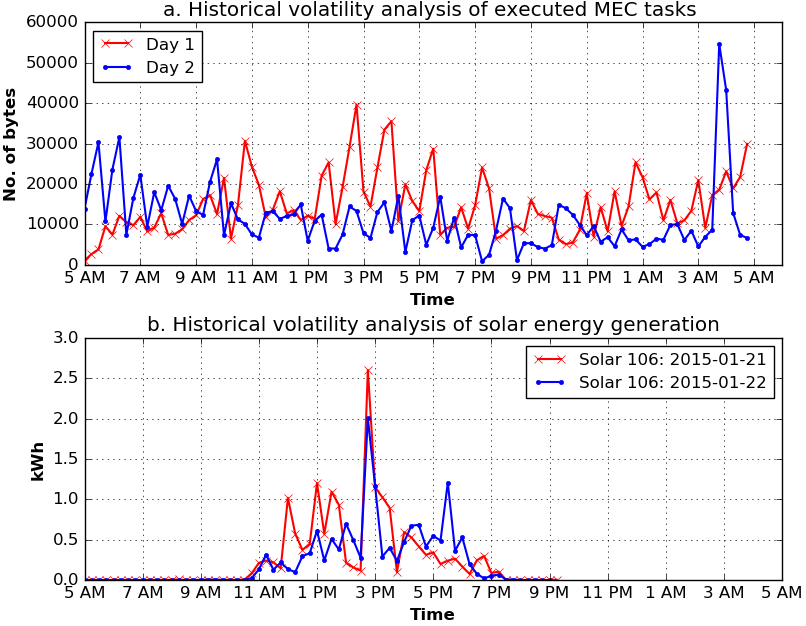}
	\caption{A historical volatility \cite{IEEEhowto:Ederington_volatility} analysis of a single MEC server's tasks execution \cite{IEEEhowto:CRAWDAD_dataset_nyupoly_video} and a solar unit (renewable) energy generation \cite{IEEEhowto:UMass_Solar_panel_dataset} for two consecutive days.}
	\label{fig:volatility}
		\vspace{-4mm}
\end{figure}

In this paper, we consider a grid-connected and dedicated microgrid that supports MEC-enabled wireless networks to reduce the usage of brown (i.e., non-renewable) energy for the considered network infrastructure. Therefore, the microgrid is always connected to the power grid, and also it can always be bought from the power grid to guarantee seamless energy flow to the MEC network. However, it is imperative to tackle the challenge of when to buy and how much energy to be bought from the main grid. Further, it is also essential to take an autonomous energy decision (i.e., store/buying) for the microgrid-enabled MEC network to reduce the risk of energy shortfall (i.e., the gap between demand and supply estimation).
The request of uncertain tasks to MEC compels a volatile energy consumption, while the randomness of renewable energy generation admits the volatility in a microgrid (as seen in Fig. \ref{fig:volatility}). 
In figures 1(a) and 1(b), we explain the nature of MEC tasks execution request and the energy generation of a solar unit for the consecutive day. In the figures 1(a) and 1(b), we observe that a sudden increase of the MEC tasks execution request can demand more energy to fulfill its user task, while a sharp decrease of solar generation causes energy generation volatility. In particular, Fig. 1(a) shows that the MEC tasks execution request becomes more than double between $3$ AM to $5$ AM compared to the following day. To this end, historical volatility \cite{IEEEhowto:Ederington_volatility} of a single MEC server's tasks request \cite{IEEEhowto:CRAWDAD_dataset_nyupoly_video} and a solar unit's (renewable) energy generation \cite{IEEEhowto:UMass_Solar_panel_dataset} infer (in Fig. 1) that the volatility of energy demand-supply induces a risk toward an efficient energy scheduling of MEC network.
The volatility \cite{IEEEhowto:Ederington_volatility} of energy consumption-generation possibly induces a shortfall between scheduled energy demand and supply.
This characteristic of energy demand-supply raises a risk towards an efficient energy scheduling of the microgrid-powered MEC network, where the volatility is directly proportional to the risk. 
Thus, to mitigate this challenge, risk-aware energy scheduling in which the uncertainties of both MEC energy consumption and microgrid generation are considered is essential. 
In the literature, conditional value-at-risk (CVaR) \cite{IEEEhowto:Chow_CVaR,IEEEhowto:Alsenwi_CVAR} is an effective metric to estimate the expected shortfall in risk measurement for establishing seamless energy management.
In particular, the CVaR can quantify the amount of tail risk (i.e., both positive and negative), which is used to characterize the expected residual of energy demand and supply.
To prevent the energy shortfall between consumption and generation estimation, a risk-aware energy scheduling mechanism is needed to establish an efficient control of the non-renewable energy generation (i.e., turn on/off non-renewable energy generation) at the main grid for the considered microgrid-powered MEC network.
Throughout this paper, the term \emph{energy shortfall} represents the energy residual between demand and supply (i.e., consumption and generation).
One of this work's core objectives is to maximize the renewable energy usage for the microgrid-powered MEC network. Subsequently, the proposed approach can efficiently take an autonomous control decision of the energy store or buy for the considered network. When renewable energy generation is more than the demand, the proposed system can trigger the energy store decision for storing a surplus amount of energy into an energy storage medium for future usage. In contrast, the microgrid-powered MEC network buys non-renewable energy from the main-grid. However, the advance storing of non-renewable energy into an energy storage medium or always buying energy from the main grid is conflicting with the principle of the microgrid \cite{IEEEhowto:MicroGrids_Lasseter, IEEEhowto:MicroGrids_deployment_Rinaldi}. Therefore, in this work, we cannot keep non-renewable energy in the storage medium for future usage, nor we can continuously buy from the power grid.

We summarize the goal and technical motivations of our work as follows:
\begin{itemize}
	\item The goal of this work is to provide an efficient energy scheduling scheme for a microgrid-powered MEC network. Thus, the objective is to minimize the expected energy residual between estimated demand and supply by capturing the tail-risk of the uncertain energy generation and demand. In particular, the objective is to reduce the gap between energy generation and demand estimation, where it can maximize the usage of renewable energy by satisfying its constraints. As a result, the proposed scheduling mechanism can minimize the amount of energy to be bought from the power grid since the usage of renewable energy becomes maximized. Further, this proactive scheduling can protect the MEC network from the risk of energy outage or such an emergency situation by applying autonomous control decisions (i.e., store/buying action) by the microgrid controller. 
	\item To achieve the above goal, we need to design the solution by considering the following aspects: 
		1) uncertainty for both energy consumption and generation should be taken into account, 
		2) coordinating between wireless network energy consumption and microgrid generation, and  
		3) characterizing strong temporal dependencies among the massive amounts of multidimensional energy data (i.e., the energy consumption of the MEC network and generation of microgrid). Therefore, capturing the dynamic of a Markov decision process can be one of the possible ways to overcome those technical challenges. Thus, a model-free, policy-based, and on-policy reinforcement learning mechanism can be a suitable way since it does not build any model, directly approximates the energy scheduling policy, and require new energy data (i.e., new environment) to obtain the policy. In general \cite{IEEEhowto:He_Green_RL, IEEEhowto:softmax_Sutton}, single-agent mechanism is not capable of choosing an action for the best policy due to limited information and the single-agent reinforcement learning only optimizes the action policy for itself only. However, by changing the environment this method cannot cope with an unknown environment due to diversity. Instead, in a multi-agent system, each agent learns action policy based on diverse feedbacks from other agents or global agent and optimizes the decision towards the best energy scheduling.   
\end{itemize}
\subsection{Challenges}
To design a \textit{risk-aware energy scheduling} for MEC, we face several challenges:
\begin{itemize}
	\item First, the volatilities of both energy consumption and generation induce risk of energy shortfall to the MEC network. Inefficient energy estimation for both demand and supply result in a high risk of energy demand-supply failures, thus increasing the power outage probability. Hence, proper coordination between the energy consumption of a MEC network and the energy supply of the microgrid can reduce the risk of energy shortfall between demand and supply.
	
	\item Second, how to coordinate between wireless network energy consumption and microgrid generation for balancing energy demand-supply is critical. Long-term energy consumption and generation estimation from historical energy data (i.e., energy consumption of the MEC network and generation of microgrid) can impose an efficient energy scheduling, which can help to reduce the risk of energy shortfall and overcomes the uncertainties for advanced demand-supply management. However, the energy data not only holds the multidimensionality but also establishes a strong temporal dependence (autocorrelation) among them.    
	
	\item Third, how to characterize the massive amount of multidimensional energy data that can capture the autocorrelation representing a time variant information of demand-supply. Here, the dynamics that can be assumed as a Markovian model so that a multi-agent model can be one possible way to overcome this challenge.
\end{itemize}
\subsection{Contribution}
To alleviate from the above challenges, we focus on approaches that not only consider risk measurement in both energy consumption and generation, but also provide efficient energy scheduling for the MEC network such that demand is satisfied. We summarize our key contributions as follows: 
\begin{itemize}
	\item First, we formulate a \emph{risk-aware energy scheduling} problem for the microgrid-powered MEC network, where the objective is to reduce the loss of energy demand-supply estimation such that the CVaR confidence level is satisfied. This optimization problem not only coordinates between MEC energy consumption and microgrid generation, but also considers the expected shortfall in CVaR risk measurement with a long tail distribution, where we show that the formulated problem is NP-hard.  

    \item Second, to achieve optimal scheduling of the formulated problem, we analyze and remodel it with multi-agent deep reinforcement learning (MADRL), which is similar to an $N$-agent stochastic game with a joint policy. We show that this game not only guarantees at least one Nash equilibrium point in stationary strategies, but also ensures the convergence of the proposed model. 
	
	\item Third, to derive the solution of proposed model, we apply a MADRL-based asynchronous advantage actor-critic (A3C) algorithm and achieve the optimal policy, in which we employ shared neural networks (weight sharing) for low computational complexity. This MADRL model overcomes the curse of dimensionality for the state-space and accepts the best policy among the other agents toward updating the global policy with less information.
	
	\item Finally, we perform an extensive experimental analysis for the proposed risk-aware energy scheduling model, and the experimental results show the proposed approach outperforms the single agent A2C and random-agent A3C solution in terms of energy scheduling and risk measurement, achieving around $92\%$, $96\%$, and $92\%$ test accuracy with $4.72\%, 5.65\%,$ and $7.46\%$ CVaR for confidence levels of $90\%, 95\%,$ and $99\%$, respectively.  
\end{itemize}

This paper is an extension of our conference version \cite{IEEEhowto:Munir_GC_Multi_Agent} that provides one of the first models for risk-aware energy scheduling of a microgrid-powered MEC network. Apart from the previous work, in this paper, we focus to provide not only the rigorous experimental analysis but also a concrete theoretical perspective to establish the efficiency of the proposed model. In particular, we have formulated an $N$-agent stochastic game for the solution that ensures a joint policy nash equilibrium with convergence guarantee. Further, in the problem formulation, we have added three additional constraints that establish a concrete model. To this end, we characterize the tail-risk of the energy shortfall by comparing with VaR and CVaR that consider both normal and Student's t distribution \cite{IEEEhowto:Roth_heavy_tailed_process_t_distribution}. 

The rest of the paper is organized as follows. We present some related work of current research in Section II. In Section III, we describe the network energy consumption model, microgrid energy generation model, risk assessment model, and problem formulation of the risk-aware energy scheduling problem. We present risk-aware energy scheduling via an $N$-agent stochastic game and MADRL in Section IV. In Section V, we provide the experimental analysis and discussion regarding the proposed solution. Finally, we conclude the paper in Section VI.

\section{Related Work}
In this section, we discuss background of MEC and microgrid, some of the related works, and challenges, which are grouped into four categories: (i) background, (ii) MEC networks with renewable energy, (iii) sustainable demand response (DR) management, and (iv) reinforcement learning in wireless networks.
\subsection{Background}
The goal of the MEC technology is to provide computational facilities at the edge of a network by employing low-latency, high-bandwidth communication with real-time feedback for Internet of Things (IoT) applications and services \cite{IEEEhowto:Porambage_Survey_MA_IOT}. Initially, mobile edge computing was referred to as the network edge of a mobile network \cite{IEEEhowto:Mao_A_Survey_on_Mobile_Edge}. However, the MEC function is not only limited to mobile networks but also reflects on non-cellular operators' requirements. Therefore, the European Telecommunications Standards Institute (ETSI) Industry Specification Group (ISG) officially changed its name from Mobile Edge Computing ISG to Multi-Access Edge Computing ISG \cite{IEEEhowto:ETSI_STANDARD_Name}. MEC hosts are deployed in a large geographical area such as, at the edge or central data network, where User Plane Function (UPF) manages the user traffic for the targeted MEC applications in the data network \cite{IEEEhowto:ETSI_White_deployment}. Further, network operators can be responsible for the physical location of the data networks based on the supported applications, available site facilities, and requirements, where they consider technical and business parameters \cite{IEEEhowto:ETSI_White_deployment}. Nowadays, the necessity of MEC becomes a promising technology to facilitate the smart city, smart factory, smart home, and other smart infrastructures \cite{IEEEhowto:Porambage_Survey_MA_IOT, IEEEhowto:Munir_Edge_AI}. Meanwhile, microgrid has shown its competence in those smart infrastructures as an energy supplier \cite{IEEEhowto:Microgrid_Hernandez,IEEEhowto:Microgrid_Calvillo, IEEEhowto:Piovesan_microgrid_def}. Thus, microgrid is capable of supplying energy from its renewable, non-renewable, and storage energy sources (local sources) and also able to sell/buy energy to/from main grid \cite{IEEEhowto:Munir_Edge_Microgrid, IEEEhowto:Zhang_time_slot, IEEEhowto:time_6hr_microgrid}. As a result, microgrid is a suitable candidate to enable sustainable edge computing in the era of next-generation networks.

\subsection{MEC Networks with Renewable Energy}
MEC emerges via computational requirements at the edge of wireless networks to handle the massive expansion of IoT applications and services. The amount of computational data will reach $49$ exabytes per month by $2021$ and MEC will be responsible for computing $63\%$ of the total computational data \cite{IEEEhowto:Cisco_White_Paper}. Therefore, to operate sustainable MEC function, a microgrid enables facilitation of energy consumption from renewable, non-renewable, and storage sources. Sustainable energy management for renewable energy enabled wireless networks has been a focus of recent years, facilitating seamless edge computing. A study on renewable energy powered base stations (BSs) operation was performed a decade ago in \cite{IEEEhowto:White_Ericsson}, where authors have analyzed the usefulness and effectiveness of usages of renewable energy in the wireless network. Afterward, in \cite{IEEEhowto:Mao_Energyharvestingsmallcell}, small cell networks have facilitated with renewable energy, in which they investigate a network deployment methodology and design the network operation function. Recently, a learning-based scheme has proposed in \cite{IEEEhowto:Wei_RL_Scheduling_Resource}, where user scheduling and network resource allocation have performed of heterogeneous networks by considering hybrid energy supply. To facilitate sustainable MEC, it is essential to study the energy demand for MEC computational tasks as well as the energy generated from renewable energy sources. 

In recent years, MEC has faced challenges related to low latency scheduling, scalability, and sustainability, in which proper energy scheduling can solve the sustainability of MEC due to energy failure \cite{IEEEhowto:Shahzadi_Multi_access, IEEEhowto:Li_Egde_Re}. To promote energy saving for user devices, energy-efficient task offloading \cite{IEEEhowto:Chen_mobile_edge_computing} and dynamic task offloading are considered in \cite{IEEEhowto:Huang_Offloading}, increasing the energy consumption of MEC networks. Therefore, to enable smart grid powered mobile networks, a joint method was proposed for both BS operation and power distribution, which provides a strong relationship between BS operation power consumption and smart grid power generation \cite{IEEEhowto:Huang_Jointly_Optimizing_BS}. A hybrid energy supply aware user scheduling and resource allocation scheme has developed for HetNets environment, where every small cell base station (SBS) is enabled with renewable energy sources \cite{IEEEhowto:Wei_RL_Scheduling_Resource}. Further, a wired energy transferred (via energy harvesting and energy load balancing) mechanism for the BSs has been established, with a grid architecture that considers task offloading decisions among the BSs \cite{IEEEhowto:Huang_Energy_sharin}. In \cite{IEEEhowto:Hong_wireless_SmartGrids}, a method for energy-efficient wireless data transmission has been proposed that minimizes the power loss for both power generation and wireless network power consumption. However, these studies have overlooked MEC energy consumption with respect to computational tasks. Thus, a computational tasks consume $33\%$ of energy from the total consumed energy of the wireless networks \cite{IEEEhowto:Dayarathna_Survey}.

\subsection{Sustainable Demand Response Management }
A sustainability issue arises for MEC networks, where the computations of MEC are directly proportional to energy consumption \cite{IEEEhowto:Porambage_Survey_MA_IOT}. The energy consumption of MEC network is nondeterministic in nature \cite{IEEEhowto:Munir_Edge_Microgrid}. Therefore, efficient energy supply can fulfill the requirements of sustainable MEC operation. The reliability and stability of the MEC network not only depend on MEC energy consumption, but also on energy generation of renewable (e.g., solar, wind, biofuels, etc.) and non-renewable (e.g., diesel generator, coal power, and so on) sources \cite{IEEEhowto:Li_Towards_sustainable_in_situ}. A microgrid can manage those challenges by providing proper demand response (DR) management, in which a microgrid controller is an essential component to do such \cite{IEEEhowto:Li_Multiobjective_Demand}. Competence of the proper DR management is already shown in the field of residential \cite{IEEEhowto:Kong_LSTM_Home_grid,IEEEhowto:Mengistu_Home_App_Load}, commercial \cite{IEEEhowto:Siano_Energy_market}, and cloud data center \cite{ IEEEhowto:Nguyen_Incentivizing_DR_DataCenter,IEEEhowto:Pham_DR} energy management. In  \cite{IEEEhowto:Xu_Risk_Management}, a risk management scheme was introduced for optimizing the midterm power portfolio in energy market, where reliability of energy pricing is increased since risk measurement has included. Volatility of energy demand and supply induces a risk of energy failure, indicating the significance of energy scheduling of the microgrid-powered MEC networks. As a result, to ensure a sustainable energy scheduling, we discretize the risk under a CVaR \cite{IEEEhowto:Chow_CVaR, IEEEhowto:Alsenwi_CVAR}, where the risk of energy shortfall is characterized by a long-tail distribution.

\subsection{Reinforcement Learning in Wireless Network}
A variety of reinforcement learning (RL) approaches have been used to solve complex problems such as user tasks offloading, software-defined network (SDN) management, strategic planning of energy markets, and SBS network resources allocation \cite{IEEEhowto:Alam_Reinforcement,IEEEhowto:Munir_QL, IEEEhowto:Rahimiyan_Q_learning, IEEEhowto:Bairagi_Q_learning_Res_Alo_SBS, IEEEhowto:Wei_RL_Scheduling_Resource,IEEEhowto:He_Green_RL}. Multi-agent reinforcement learning is used to solve spatio-temporal resource assignment problems \cite{IEEEhowto:Alam_Reinforcement}. An $\epsilon$-greed-based RL method has been applied to SDN-based smart city service management \cite{IEEEhowto:Munir_QL}, and a single agent Q-learning model has studied for energy market analysis in \cite{IEEEhowto:Rahimiyan_Q_learning}. Furthermore, deep Q-networks (DQN) have been employed for green resource management in content-centric IoT networks \cite{IEEEhowto:He_Green_RL}. A DQN-based single agent actor-critic model has been proposed for task scheduling in renewable energy powered heterogeneous networks \cite{IEEEhowto:Wei_RL_Scheduling_Resource}, and a data-driven model has developed for the BS energy saving mechanism in \cite{IEEEhowto:Li_DeepNapu}. However, these approaches are very expensive with respect to computation, whereas multi-agent deep reinforcement learning (MADRL) with shared neural networks is one possible way to obtain a low computational complexity solution. In order to solve the risk-aware energy scheduling problem with optimal results and fast convergence, a MADRL-based asynchronous A3C model is more appropriate \cite{IEEEhowto:Lowe_Multi_agent_actor_critic, IEEEhowto:Mnih_async_a3c}. 

In this paper, we tackle the risk-aware energy scheduling problem for the microgrid-powered MEC networks by considering both energy consumption and generation under risk measurement. The detail discussion of the system model and problem formulation are given in the following section.

\section{System Model and Problem Formulation}
\begin{figure}[!t]
	\centerline{\includegraphics[width=8.9cm]{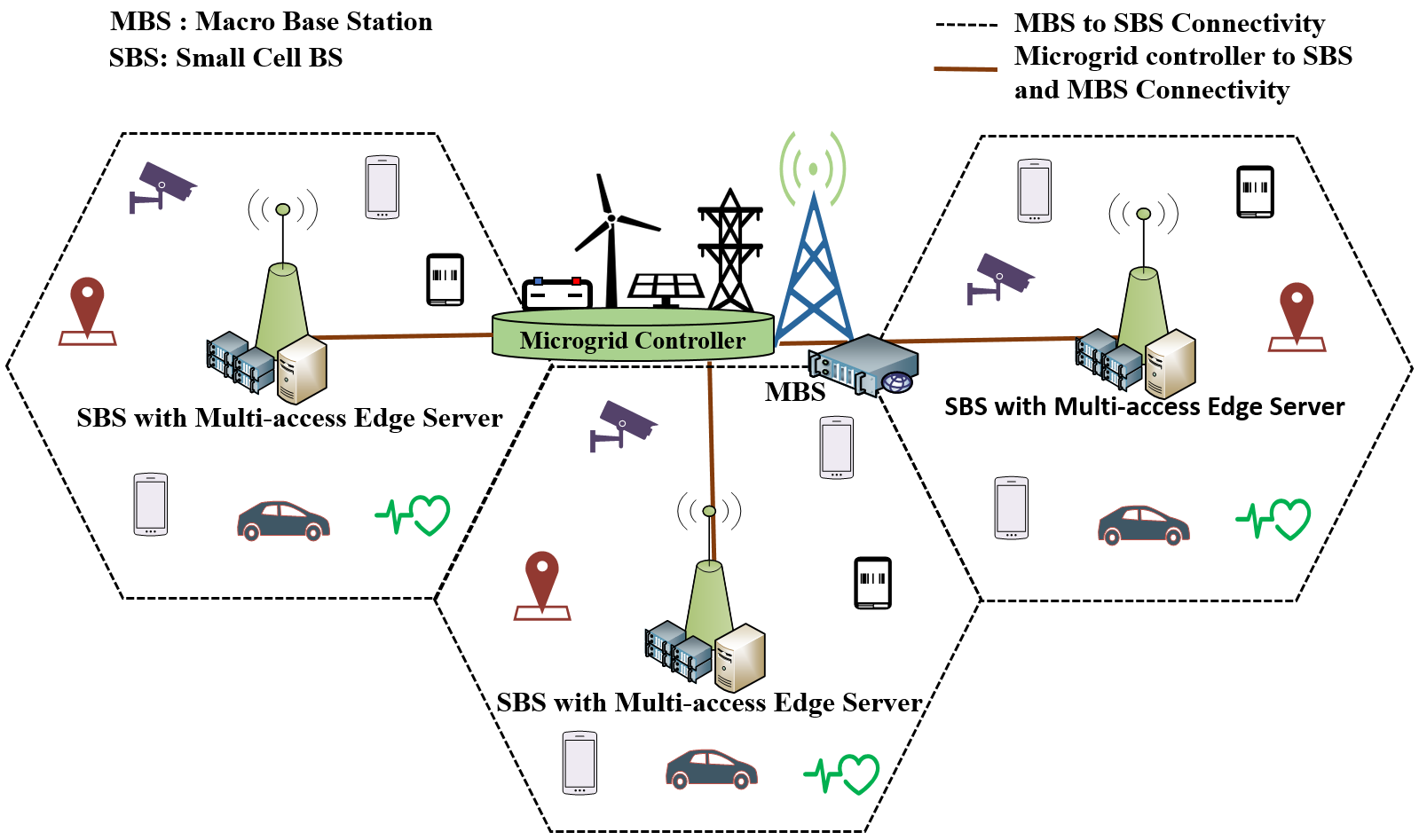}}
	\caption{System model for edge computing with microgrid.}
	\label{System_model}
	\vspace{-4mm}
\end{figure}
The multi-access edge servers are capable of performing heterogeneous computational tasks (e.g., smart health, VR, emergency monitoring and so on) and these tasks are associated with SBSs as shown in Fig. \ref{System_model}. The energy demand of this MEC network is fulfilled by the microgrid energy sources (e.g., renewable, non-renewable, and storage), and a microgrid controller can communicate with the macro base station (MBS) for managing the energy supply based on MEC network energy demand. All the SBSs are physically connected with the MBS and controlled by the same network operator. Here, the risk of energy failure is a fundamental challenge for sustainable edge computing \cite{IEEEhowto:Shahzadi_Multi_access} in existing microgrid-powered MEC networks, where the energy consumption of edge computing is random over time and renewable energy generation is nondeterministic in nature \cite{IEEEhowto:Li_Egde_Re, IEEEhowto:Munir_Edge_Microgrid}. Therefore, the microgrid-powered MEC network system model can be decomposed into three parts, the network energy consumption model, microgrid energy generation model, and risk assessment model, which are discussed in the later sub-sections in detail.
\begin{table}[t!]
	\caption{Summary of Notation}
	\begin{center}
		\begin{tabular}{|c|c|}
			\hline
			\textbf{Notation}&{\textbf{Description}} \\
			\hline
			$\mathcal{B}$&Set of SBSs under the MBS\\
			\hline
			$\mathcal{C}_i$&Set of active server under SBS $i \in \mathcal{B}$\\
			\hline
			$\lambda_i (t)$&Task arrival rate at SBS $i \in \mathcal{B}$\\
			\hline
			$\chi_i(t)$&Average traffic size at SBS $i \in \mathcal{B}$\\
			\hline
			$\varphi_i(t)$& Data rate at SBS $i \in \mathcal{B}$\\
			\hline
			$\mu_i(t)$&Service rate at SBS $i \in \mathcal{B}$\\
			\hline
			$\rho_i(t)$&Server utilization at SBS $i \in \mathcal{B}$\\
			\hline
			$\mathcal{E}^{\textrm{dem}}(t)$& Total energy demand for $\forall i \in \mathcal{B}$\\
			\hline				
			${\mathcal{E}}^{\textrm{ren}}(t)$& Renewable energy generation at time slot $t$ \\
			\hline
			${\mathcal{E}}^{\textrm{non}}(t)$& Non-renewable energy generation at time slot $t$ \\
			\hline
			${\mathcal{E}}^{\textrm{sto}}(t)$& Stored energy at time slot $t$ \\
			\hline
			${\mathcal{E}}^{\textrm{gen}}(t)$& Total energy generation at time slot $t$ \\
			\hline	
			$\mathcal{N}$& Set of virtual agents\\
			\hline
		\end{tabular}
		\label{tab1_SummaryofNotation}
	\end{center}
		\vspace{-6mm}
\end{table}
\subsection{Network Energy Consumption Model}
In the network energy consumption model, we consider a set of SBSs $\mathcal{B} = \left\{{1,2,\dots, B}\right\}$ that are deployed under a macro base station (MBS), where each SBS $i$ has a set of MEC heterogeneous active servers $\mathcal{C}_i = \left\{{1,2,\dots,C_i}\right\}$ and these MEC servers are physically deployed with SBS $i$ \cite{IEEEhowto:ETSI_White_Paper}. We consider one energy consumption cycle as a set $\mathcal{T} = \left\{{1,2,\dots, T}\right\}$ of a finite time horizon, where the length of a discrete time slot $t$ is a 15-minute duration (i.e., one observational period), in general, the total duration of each cycle is one month \cite{IEEEhowto:Zhang_time_slot}. In the typical wireless networks \cite{IEEEhowto:Borst_network_time, IEEEhowto:Han_time_10_min}, the per-user channel and traffic variation period is up to $0.1$ second and traffic aggregation demand variation is $0.1$ minute. In fact, we can predict the user association and network status variations by the duration of several minutes, hours, and days \cite{IEEEhowto:Huang_Jointly_Optimizing_BS,IEEEhowto:Borst_network_time,  IEEEhowto:Han_time_10_min}. On the other hand, the microgrid (i.e., renewable, non-renewable, and storage) energy generation and supply model belongs to several time slot durations correspondingly up to $15$ minutes, $1$ hours, $6$ hours, $1$ day, several days and so on \cite{IEEEhowto:Zhang_time_slot,IEEEhowto:Microgrid_Hernandez,IEEEhowto:UMass_Solar_panel_dataset,IEEEhowto:Munir_Edge_Microgrid,IEEEhowto:time_6hr_microgrid}. The observation of a time slot $t$ is ended at the $15$-th minute and capable of capturing the changes of network dynamics for entire $15$ minutes duration \footnote{Consider fast-changing channel condition (e.g., variation period of $0.1$ second), the $15$ minutes observational period is comprised of all changes \\$\left\{{0.1, 0.2, 0.3,\dots 900}\right\}$ (in second). In fact, if we consider $10$ ms, the dynamic of the changes also be captured during each $15$ minutes period.}. Therefore, $15$ minutes time slot is reasonable for the proposed scenario. Let us consider a smart city scenario, where the SBS $i$ can serve a set of heterogeneous user tasks $\mathcal{K} = \left\{{1,2,\dots,K}\right\}$ (e.g., video surveillance, emergency health care, smart transportation and so on.) at time slot $t$. The user association between SBS $i \in \mathcal{B}$ and task $k\in \mathcal{K}$ is denoted as $\Omega_{ik}(t)$. We assume $\Omega_{ik}(t) = 1$ if task $k$ is assigned to SBS $i$ at time $t$, and $0$ otherwise. Moreover, SBSs are considered as up and running to anticipate those user requests. For example, we have two SBS $i_1$ and $i_2$ in the microgrid powered MEC network. At the beginning of an observational time period of $0$ to $15$ minutes (i.e., indexed as $t = 1$), user tasks $k_{11}, k_{12}$ and $k_{13}$ are associated with the SBS $i_1$ by the associations $\Omega_{11}(t)=1$, $\Omega_{12}(t) =1$, and $\Omega_{13}(t) = 1$, respectively. Similarly, user tasks $k_{21}$ and $k_{22}$ are associated with the SBS $i_2$ with the association $\Omega_{21}(t)=1$, and $\Omega_{22}(t) =1$, respectively. After the $2$ minutes during of the total $15$ minutes observation period, the user task $k_{11}$ of the SBS $i_1$ is moved to the SBS $i_2$. As a result, a new user task $k_{23}$ will assign to SBS $i_2$ by $\Omega_{23}(t)=1$. Meanwhile, the association status of user task $k_{11}$ at SBS $i_1$ will be changed from $\Omega_{11}(t)=1$ to $ \Omega_{11}(t)=0$. During the one observational period these changes are \emph{known to microgrid controller} for the microgrid powered MEC network. The microgrid controller calculates the energy consumption of first $2$ minutes for that user task with SBS $i_1$, and rest of the duration with SBS $i_2$. Therefore, if the users are moving fast that do not affect the total energy consumption calculation of the considered scenario. The main notations that are used in this work is represented in Table \ref{tab1_SummaryofNotation}.

\subsubsection{Network Operation Energy Consumption}
We consider a task arrival rate $\lambda_i (t)$ for SBS $i$ with an average traffic size $\chi_i(t)$ at time slot $t$, where the task arrival rate follows a Poisson process and $\lambda_i (t) \chi_i(t)$ is the average traffic load. The capacity of SBS $i \in \mathcal{B}$ to receive the user tasks $\forall k \in \mathcal{K}$ is as follows \cite{IEEEhowto:Sun_EMM,  IEEEhowto:Munir_Edge_Microgrid, IEEEhowto:Abedin_Fog_Load_Balancing}:
\begin{equation} \label{eq:data_rate}
\varphi_{i}(t) = \sum_{k \in \mathcal{K}} \Omega_{ik}(t) w_{ik} \log_2\Big(1 + \frac{p_{ik} g_{ik}(t)}{\sigma^2_{ik} + \sum_{j \in \mathcal{B}, j \ne i}I_{j}(t)}\Big), \forall i \in \mathcal{B}, 
\end{equation}
where $w_{ik}$ is the fixed channel bandwidth assigned to $k \in \mathcal{K}$, $p_{ik}$ is the transmission power between task $k \in \mathcal{K}$ and SBS $i \in \mathcal{B}$, $g_{ik}(t)$ determines the channel gain, the variance of the Additive white Gaussian noise (AWGN) is denoted by $\sigma^2_{ik}$, and $I_{j}(t)$ is the channel interference with other SBSs. Therefore, the average service rate for SBS $i$ is calculated as follows:
\begin{equation} \label{eq:mg1_service_rate}
\mu_i(t) = \frac{\varphi_{i}(t)}{\chi_i(t)}.
\end{equation}
The SBS $i\in \mathcal{B}$ data rate is considered as constant and the traffic size is already known, so the service rate follows an exponential distribution. As a result, the M/M/1 queuing model can be considered as an appropriate choice \cite{IEEEhowto:Munir_Edge_Microgrid,IEEEhowto:Chen_MM1,IEEEhowto:Abedin_MM1_Fog}. At time slot $t$, tasks $\forall k \in \mathcal{K}$ are uniformly distributed under SBS $i$ and the overall server utilization rate is as follows \cite{IEEEhowto:Huang_Jointly_Optimizing_BS}:
\begin{equation} \label{eq:server_utilization}
\rho_i(t) = \sum_{k \in \mathcal{K}} \Omega_{ik}(t) \frac{\lambda_i(t) }{\mu_i(t)},
\end{equation}
where $\sum_{k \in \mathcal{K}} \Omega_{ik}(t) \lambda_i(t)$ is the total amount of served tasks at time slot $t$ by SBS $i$. 

The energy consumption of each SBS $i \in \mathcal{B}$ is comprised of two types of energy consumptions: 1) SBS general operation (i.e., up and running), and 2) data transfer through the network (i.e., payload communications). In particular, the general operation energy consumption is considered as a static energy $\mathcal{E}^{\textrm{net}}_{\textrm{st}}(t)$ that is the energy consumption  of each SBS $i \in \mathcal{B}$ without carrying any traffic load (i.e., meta-data and payload) at time slot $t$ \cite{IEEEhowto:Han_grid_sdn, IEEEhowto:Huang_Jointly_Optimizing_BS, IEEEhowto:Munir_Edge_Microgrid, IEEEhowto:Han_time_10_min, IEEEhowto:Total_Energy_Auer}.  Further, the static energy consumption differs among the SBSs based on its configurations and capacity\cite{IEEEhowto:Total_Energy_Auer} while this energy is fixed during the observational period of $t$.  Meanwhile, the payload communication energy consumption is taken into account as a dynamic energy consumption of each SBS $i \in \mathcal{B}$ \cite{IEEEhowto:Huang_Jointly_Optimizing_BS,IEEEhowto:Total_Energy_Auer}. In particular,  dynamic energy consumption relies on the network traffic communications that is the traffic load of each SBS $i \in \mathcal{B}$ at time slot $t$. In fact, the dynamic energy consumption not only depends on traffic size but also relies on SBSs types \cite{IEEEhowto:Total_Energy_Auer}. Thus, a linear model is suitable to estimate the energy consumption of each SBS $i \in \mathcal{B}$ at time slot $t$ and the network operation energy consumption is determined as follows \cite{IEEEhowto:Han_grid_sdn, IEEEhowto:Huang_Jointly_Optimizing_BS, IEEEhowto:Han_time_10_min, IEEEhowto:Total_Energy_Auer}:
\begin{equation} \label{eq:comm_energy}
\mathcal{E}^{\textrm{net}}_i(t) =  \eta^{\textrm{net}}(t)\rho_i(t)  + \mathcal{E}^{\textrm{net}}_{\textrm{st}}(t),
\end{equation}
where $\eta^{\textrm{net}}$ represents the energy coefficient and the value of parameter $\eta^{\textrm{net}}(t)$ depends on physical components of SBS as well as transfered payload size \cite{IEEEhowto:Huang_Jointly_Optimizing_BS, IEEEhowto:Total_Energy_Auer}.

\subsubsection{MEC Server Computational Energy}
MEC server energy consumption depends on the number of CPU cores, activity ratio, and processor architecture. Each MEC server consists of $L$ number of homogeneous CPU cores with $M$ numbers of CPU component (i.e., FE: frontend, INT: integer units, FP: floating point units, BPU: branch prediction unit, L1: L1 cache, L2: L2 cache,
and MEM: FSB and main memory.) \cite{IEEEhowto:Bertran_Multi_Core_CPU}. Therefore, the dynamic energy consumption for core $l$ with $M$ components is determined by $\sum_{m \in M}\eta^{\textrm{cpu}}_m\delta_{ml}(t)$, where $\eta^{\textrm{cpu}}_m$ is the weight of component $m$ and $\delta_{ml}(t)$ represents the activity ratio at time slot $t$ and $ \sum_{c \in \mathcal{C}_i} \sum_{l \in L} \sum_{m \in M} \delta_{ml}(t) \approx \frac{1}{\rho_i(t)}$. The general operation (e.g., signal messaging, idle state) of the MEC server consumes a static energy $\mathcal{E}^{\textrm{mec}}_{\textrm{st}}(t)$. Thus, the total energy consumption for the multi-core MEC server is defined as follows \cite{IEEEhowto:Bertran_Multi_Core_CPU}: 
\begin{equation} \label{eq:cpu_energy}
\mathcal{E}^{\textrm{mec}}_i(t) =\sum_{c \in \mathcal{C}_i} \bigg( \sum_{l \in L} \sum_{m \in M}  \eta^{\textrm{cpu}}_m\delta_{ml}(t)  + \mathcal{E}^{\textrm{mec}}_{\textrm{st}}(t)\bigg). 
\end{equation}
In \eqref{eq:cpu_energy}, since the activity ratio $\delta_{ml}(t)$ of the each core depends on the server capacity, this ensures variability of the MEC servers energy consumption. 
\subsubsection{Total Energy Consumption}
The total energy consumption of MEC enabled SBS $i \in \mathcal{B}$ includes the static energy $\mathcal{E}^{\textrm{st}}_i(t)$ and the dynamic energy $\mathcal{E}^{\textrm{dyn}}_i(t)$ \cite{IEEEhowto:Total_Energy_Auer} at time slot $t$. Equations \eqref{eq:comm_energy} and \eqref{eq:cpu_energy} can determine the energy consumption for the SBS network's operation and MEC servers usages, respectively. Thus, in both cases, the dynamic energy $\mathcal{E}^{\textrm{dyn}}_i(t)$ relies on overall utilization rate $\rho_i(t)$ (in \eqref{eq:server_utilization}) at each MEC enabled SBS $i \in \mathcal{B}$. In particular, network's operation dynamic energy consumption depends on $\rho_i(t)$ (in \eqref{eq:comm_energy}) while dynamic energy consumption of MEC servers' relies on activity ratio $\sum_{c \in \mathcal{C}_i} \sum_{l \in L} \sum_{m \in M} \delta_{ml}(t) \approx \frac{1}{\rho_i(t)}$ of those servers (in \eqref{eq:cpu_energy}) at time slot $t$. Here, the total dynamic energy consumption of each MEC enabled SBS $i \in \mathcal{B}$ for time slot $t$ is defined as follows:
\begin{equation} \label{eq:edge_dynamic_eng}
\mathcal{E}^{\textrm{dyn}}_i(t) =\eta^{\textrm{net}}(t)\rho_i(t)  + \frac{1}{\rho_i(t)} \sum_{c \in \mathcal{C}_i} \sum_{l \in L} \sum_{m \in M}  \eta^{\textrm{cpu}}_m.
\end{equation}
Further, we can denote the total amount of static energy consumption of each MEC enabled SBS $i \in \mathcal{B}$ for time slot $t$ as $\mathcal{E}^{\textrm{st}}_i(t) = \mathcal{E}^{\textrm{net}}_{\textrm{st}}(t) + \sum_{c \in \mathcal{C}_i}  \mathcal{E}^{\textrm{mec}}_{\textrm{st}}(t)$, where $\mathcal{E}^{\textrm{net}}_{\textrm{st}}(t)$ and $\sum_{c \in \mathcal{C}_i}  \mathcal{E}^{\textrm{mec}}_{\textrm{st}}(t)$ represent the static energy consumption of network operation and MEC severs, respectively. Therefore, the total energy demand of each  MEC enabled SBS $i \in \mathcal{B}$ is as follows \cite{IEEEhowto:Total_Energy_Auer}:
\begin{equation} \label{eq:edge_total_eng}
\mathcal{E}^{\textrm{dem}}_i(t) = \eta^{\textrm{net}}_{i}(t)\rho_i(t)  + \frac{1}{\rho_i(t)} \sum_{c \in \mathcal{C}_i} \sum_{l \in L} \sum_{m \in M}  \eta^{\textrm{cpu}}_m + \mathcal{E}^{\textrm{st}}_i(t),
\end{equation}
where $\eta^{\textrm{net}}_{i}(t)$ represents the energy coefficient of dynamic energy consumption for the network operation at SBS $i$ \cite{IEEEhowto:Total_Energy_Auer} while $\eta^{\textrm{cpu}}_m$ is the weight of each CPU component $m$ for core $l$ at MEC server $c \in \mathcal{C}_i$ \cite{IEEEhowto:Bertran_Multi_Core_CPU}. Here, $\mathcal{E}^{\textrm{st}}_i(t)$ denotes the overall static energy consumption of each MEC enabled SBS $i \in \mathcal{B}$ at time slot $t$. Therefore, the total energy demand for $B$ SBSs that are encompassed to the MBS is as follows:
\begin{equation} \label{eq:All_BS_total_eng}
\mathcal{E}^{\textrm{dem}}(t) = \sum_{ i \in \mathcal{B}} \big(\eta^{\textrm{net}}_{i}(t)\rho_i(t)  + \frac{1}{\rho_i(t)} \sum_{c \in \mathcal{C}_i} \sum_{l \in L} \sum_{m \in M}  \eta^{\textrm{cpu}}_m + \mathcal{E}^{\textrm{st}}_i(t)\big).
\end{equation}
In \eqref{eq:All_BS_total_eng}, $\eta^{\textrm{net}}_{i}(t)$ is network operation energy consumption coefficient of each SBS $i \in \mathcal{B}$, where \eqref{eq:All_BS_total_eng} is a univariate (single-variable) quadratic function. The value of $\eta^{\textrm{net}}_{i}(t)$ for each SBS $i \in \mathcal{B}$ is fixed and relies on types of BSs (i.e.,  number of transceiver and configuration) \cite{IEEEhowto:Total_Energy_Auer}. We consider each SBS $i$ as a Micro base station with two radio head transceiver, where the value of the coefficient parameter is $\eta^{\textrm{net}}_{i}(t) = 2.6$ \cite{IEEEhowto:Total_Energy_Auer}.

\subsection{Microgrid Energy Generation Model}
The microgrid can supply both renewable energy (e.g., solar, wind, biofuels, etc.) and non-renewable energy (e.g., diesel generator, coal power, and so on). Here, ${\mathcal{E}}^{\textrm{ren}}(t)$ and ${\mathcal{E}}^{\textrm{non}}(t)$ denote the amount of renewable energy generation and non-renewable energy generation at time slot $t$, respectively. Let ${\mathcal{E}}^{\textrm{gen}}_{\textrm{max}}(t)$ denote the maximum energy generation capacity and ${\mathcal{E}}^{\textrm{gen}}(t) \le {\mathcal{E}}^{\textrm{gen}}_{\textrm{max}}(t)$ denote total amount of energy generation. The total energy generation ${\mathcal{E}}^{\textrm{gen}}(t)$ includes both renewable and non-renewable generated energy at time slot $t$, which  can be calculated as follows:
\begin{equation} \label{eq:energy_generation}
	{\mathcal{E}}^{\textrm{gen}}(t) ={\mathcal{E}}^{\textrm{ren}}(t) + {\mathcal{E}}^{\textrm{non}}(t).
\end{equation} 

We consider available storage energy ${\mathcal{E}}^{\textrm{sto}}(t-1)$ at time slot $t$ that can use to fulfill energy demand $\mathcal{E}^{\textrm{dem}}(t)$ for the network. Therefore, we define the total suppliable energy for the network at time slot $t$ as, $\mathcal{E}^{\textrm{tot}}(t) = {\mathcal{E}}^{\textrm{gen}}(t) + {\mathcal{E}}^{\textrm{sto}}(t-1)$. When the energy demand $\mathcal{E}^{\textrm{dem}}(t)$ of the network is larger than the suppliable energy $\mathcal{E}^{\textrm{tot}}(t)$ (i.e., $\mathcal{E}^{\textrm{dem}}(t) > \mathcal{E}^{\textrm{tot}}(t)$), the microgrid can obtain additional energy from the main grid for meeting extra demand \cite{IEEEhowto:Munir_Edge_Microgrid,IEEEhowto:Zhang_time_slot,IEEEhowto:Xu_energy_store}.
Thus, in case of $\mathcal{E}^{\textrm{dem}}(t) > \mathcal{E}^{\textrm{tot}}(t)$, the additional amount of energy is calculated as,
\begin{equation} \label{eq:energy_buy}
{\mathcal{E}}^{\textrm{buy}}(t) =\mathcal{E}^{\textrm{dem}}(t) - {\mathcal{E}}^{\textrm{tot}}(t).
\end{equation}
Furthermore, when $\mathcal{E}^{\textrm{dem}}(t) \le \mathcal{E}^{\textrm{tot}}(t)$, the microgrid can store surplus amount of energy ${\mathcal{E}}^{\textrm{sto}}(t)$ to storage medium. With a storage capacity ${\mathcal{E}}^{\textrm{sto}}_{\textrm{max}}(t)$ \cite{IEEEhowto:time_6hr_microgrid,IEEEhowto:Nottrott_store_capacity}, the microgrid preserves an extra amount of generated energy ${\mathcal{E}}^{\textrm{sto}}(t) \le {\mathcal{E}}^{\textrm{sto}}_{\textrm{max}}(t)$ for future use and that is determined as follows:
\begin{equation} \label{eq:energy_store}
	{\mathcal{E}}^{\textrm{sto}}(t) ={\mathcal{E}}^{\textrm{gen}}(t) + {\mathcal{E}}^{\textrm{sto}}(t-1) - \mathcal{E}^{\textrm{dem}}(t).
\end{equation}
Hence, using \eqref{eq:energy_buy} and \eqref{eq:energy_store}, we can define the following binary decision variable:
\begin{equation} \label{eq:zeta_decision_variable}
a_t = 
\left\{
\begin{array}{ll}
1 ,\;\text{if ${\mathcal{E}}^{\textrm{sto}}(t) \ge {\mathcal{E}}^{\textrm{buy}}(t)$, $ \forall t \in \mathcal{T}$},\;\\
0,\;\;\;\;\;\;\;\;\;\;\text{otherwise,}
\end{array}
\right.
\end{equation}
where $a_t = 1$ if the microgrid is able to fulfill energy demand from its own sources, and $0$ otherwise.

\subsection{Risk Assessment with Conditional Value-at-Risk (CVaR)}
To effectively capture the expected energy shortfall, we consider conditional value-at-risk (CVaR) that can quantify the amount of tail risk for any kind of prediction risk over a specific time frame. In particular, CVaR  \cite{IEEEhowto:Chow_CVaR,IEEEhowto:Rockafellar_cvar_base,IEEEhowto:Krokhmal_cvar_expline} is able to quantify risk beyond value-at-risk (VaR), and further it is coherent. Moreover, CVaR is capable of checking the assumptions imposed by VaR and capture the extreme risk of forecasting over several domains. CVaR metric is widely used in different field of applications, such as wireless network resource allocation \cite{IEEEhowto:Alsenwi_CVAR}, energy trading \cite{IEEEhowto:Moazeni_CVar_Power_system}, market investment \cite{IEEEhowto:Krokhmal_cvar_expline}, and so on. In fact, in the case of continuous distributions, the CVaR can effectively discretize the entire risk. Therefore, we characterize the CVaR of energy shortfall for the considered system model in the following section.

Let us consider a $p=2$ dimensional decision (i.e., store and buy) vector ${\boldsymbol{a}}_t \in \mathbb{R}^p$ that stands for energy storage and buying decision. Therefore, a set of decision vectors ${\boldsymbol{A}}_t$ holds the available decision ${\boldsymbol{a}}_t \in {\boldsymbol{A}}_t$ at time slot $t$ and the decisions are affected by uncertainties of the energy demand $\mathcal{E}^{\textrm{dem}}(t)$, renewable energy generation ${\mathcal{E}}^{\textrm{ren}}(t)$, non-renewable generation $\mathcal{E}^{\textrm{non}}(t)$, and storage energy ${\mathcal{E}}^{\textrm{sto}}(t)$. Considering a $q=4$ dimensional random vector ${\boldsymbol{s}}_t \in \mathbb{R}^q$ (i.e., ${\boldsymbol{s}}_t \coloneqq (\mathcal{E}^{\textrm{dem}}(t), \mathcal{E}^{\textrm{ren}}(t), \mathcal{E}^{\textrm{non}}(t), \mathcal{E}^{\textrm{sto}}(t)) \in \mathbb{R}^q, {\boldsymbol{s}}_t\in\boldsymbol{S}_t$), where $\boldsymbol{s}_t$ represents state information at time slot $t$ and loss function  $\Upsilon({\boldsymbol{a}}_t,{\boldsymbol{s}}_t)$. Since the decision ${\boldsymbol{a}}_t$ is involved with the characteristics of ${\boldsymbol{s}}_t$, the energy shortfall (i.e., the gap between demand and supply) is also affected by the uncertainties of ${\boldsymbol{s}}_t$. Here, we recall the total suppliable energy at time slot $t$ as $\mathcal{E}^{\textrm{tot}}(t) = {\mathcal{E}}^{\textrm{ren}}(t) + \mathcal{E}^{\textrm{non}}(t) + {\mathcal{E}}^{\textrm{sto}}(t-1)$ and the loss function (i.e., expected residual) $\Upsilon({\boldsymbol{a}}_t,{\boldsymbol{s}}_t)$ is determined as follows:
\begin{equation} \label{eq:loss_fn}
\Upsilon(\boldsymbol{a}_t,\boldsymbol{s}_t) = \underset{\boldsymbol{a}_t\in {\boldsymbol{A}}_t } \min \mathbb{E}_{\boldsymbol{a}_{t} \sim {\boldsymbol{s}}_t} \big[\sum_{\boldsymbol{a}_t\in {\boldsymbol{A}}_t} |\mathcal{E}^{\textrm{dem}}(t) - \mathcal{E}^{\textrm{tot}}(t)| \big].
\end{equation} 
To quantify the volatilities of both energy consumption and generation by employing the CVaR risk assessment metric, we characterize energy difference between demand and supply in \eqref{eq:loss_fn}. In particular, the estimation of energy residual (i.e., surplus/additional energy) \eqref{eq:loss_fn} depends on the energy store/buying decision that raises a risk toward an efficient energy scheduling of the microgrid-powered MEC network due to the uncertainty of both demand-generation. Additionally, \eqref{eq:loss_fn} can assist to discretize the risk of energy shortfall from the volatile characteristics of both energy demand and supply (i.e., generation) by capturing absolute difference between them based on the store/buying decision. Therefore, \eqref{eq:loss_fn} not only captures the energy scheduling gap between MEC energy demand and renewable energy generation but also copes with the store/buying decision.

We characterize the risk of energy shortfall (i.e., $\Upsilon(\boldsymbol{a}_t,\boldsymbol{s}_t)$) by employing CVaR, which catches up with the tail end of energy demand $\mathcal{E}^{\textrm{dem}}(t)$ and generation $\mathcal{E}^{\textrm{tot}}(t)$ (i.e., state ${\boldsymbol{s}}_t$). For each $\boldsymbol{a}_t \in \mathbb{R}$, the energy shortfall $\Upsilon(\boldsymbol{a}_t,\boldsymbol{s}_t)$ is a random variable and the probability distribution is bounded by the distribution of $\xi$. Thus, the probability distribution of  $\Upsilon(\boldsymbol{a}_t,\boldsymbol{s}_t)$ is denoted by $\psi(\boldsymbol{a}_t, \xi)$, where $\psi(\boldsymbol{a}_t, \xi) = P\left\{\boldsymbol{s}_t\in \mathbb{R} \colon \Upsilon(\boldsymbol{a}_t,\boldsymbol{s}_t) \le \xi\right\} $. Hence, $\Upsilon(\boldsymbol{a}_t,\boldsymbol{s}_t)$ is continuous in decision $\boldsymbol{a}_t$ and measurable in $\boldsymbol{s}_t$, in which for each decision $\boldsymbol{a}_t \in \boldsymbol{A}_t$ the expectation of energy shortfall belongs to $\mathbb{E}[|\Upsilon(\boldsymbol{a}_t,\boldsymbol{s}_t)|] < \infty$ \cite{IEEEhowto:Rockafellar_cvar_base,IEEEhowto:Krokhmal_cvar_expline}. Let us consider $\hat{\xi}$ is a left limit of probability distribution $\psi(\boldsymbol{a}_t, \xi)$ at $\xi$, where $\psi(\boldsymbol{a}_t, \hat{\xi}) = P\left\{\boldsymbol{s}_t \in \mathbb{R} \colon \Upsilon(\boldsymbol{a}_t,\boldsymbol{s}_t) < \xi\right\}$. Therefore, there exits a probability atom \footnote{An atom of a probability space is a measurable set that contain positive measure for each event \cite{IEEEhowto:JOHNSON_Atom, IEEEhowto:prob_book}. In other word, values of a random variable are called atoms \cite{IEEEhowto:prob_book}. Consider an atom of a probability space is a measurable set $X$ with positive measure $P(X)$ then for each subset $Y \subseteq X$, either $P(Y) = 0$ or $P(X) = P(Y)$ \cite{IEEEhowto:JOHNSON_Atom}. In our case, the difference $\psi(\boldsymbol{a}_t, {\xi}) - \psi(\boldsymbol{a}_t, \hat{\xi}) = P\left\{\boldsymbol{s}_t \in \mathbb{R} \colon \Upsilon(\boldsymbol{a}_t,\boldsymbol{s}_t) = \xi\right\} > 0$ is positive, so that $\psi(\boldsymbol{a}_t, .)$ has a jump at ${\xi}$, where $\hat{\xi}$ denotes convergence from the left to the ${\xi}$. Therefore, we can conclude that the probability atom exists at ${\xi}$ \cite{IEEEhowto:Chow_CVaR,IEEEhowto:Rockafellar_cvar_base,IEEEhowto:Krokhmal_cvar_expline}.} in $\xi$ since $\psi(\boldsymbol{a}_t, {\xi}) - \psi(\boldsymbol{a}_t, \hat{\xi}) = P\left\{\boldsymbol{s}_t \in \mathbb{R} \colon \Upsilon(\boldsymbol{a}_t,\boldsymbol{s}_t) = \xi\right\}$ is positive. In literature \cite{IEEEhowto:Alsenwi_CVAR,IEEEhowto:Chow_CVaR,IEEEhowto:Rockafellar_cvar_base,IEEEhowto:Krokhmal_cvar_expline,IEEEhowto:Moazeni_CVar_Power_system}, the measurement of CVaR is derived from value at risk (VaR) is one of the suitable ways for our scenario, where CVaR is also capable of capturing the shortcoming of VaR by quantifying tail risk of the distribution of possible decisions. We define a confidence range $\alpha \in (0,1)$, where any specified probability range $\alpha \in (0,1)$ (e.g., empirically used $\alpha = 0.90$, $\alpha = 0.95$, or $\alpha = 0.99$ \cite{IEEEhowto:Rockafellar_cvar_base,IEEEhowto:Krokhmal_cvar_expline,IEEEhowto:Moazeni_CVar_Power_system}) and the VaR $\xi_{\alpha} (\boldsymbol{a}_t)$ is defined as follows:
\begin{definition} \textbf{(VaR $\xi_{\alpha} (\boldsymbol{a}_t)$):}
	\label{def:var}
	The risk of energy shortfall is associated with a decision of $\boldsymbol{a}_t$ such that $\xi_{\alpha} (\boldsymbol{a}_t) = \underset{\xi \in \mathbb{R}} \min \; \left\{\xi \in \mathbb{R} \colon \psi(\boldsymbol{a}_t, \xi) \ge \alpha\right\}$, where value of $\xi_{\alpha} (\boldsymbol{a}_t)$ belongs to $\xi$ since $\psi(\boldsymbol{a}_t, \xi)$ is nondecreasing and continuous in $\xi$. Thus, the value at risk $\xi_{\alpha} (\boldsymbol{a}_t)$ for decision $\boldsymbol{a}_t$ is defined as follows:  
	\begin{equation} \label{eq:var_eq}
	\xi_{\alpha} (\boldsymbol{a}_t) = \underset{\xi \in \mathbb{R}} \argmin \; \psi(\boldsymbol{a}_t, \xi) \ge \alpha, 
	\end{equation}
where $\xi_{\alpha} (\boldsymbol{a}_t)$ determines the value $\xi$ such that $\psi(\boldsymbol{a}_t, \xi) = \alpha$. 
\end{definition}
To capture the extreme risk of energy shortfall which exceeds from VaR cutoff point, we define CVaR $\Phi_{\alpha} (\boldsymbol{a}_t)$ as the expectation of worst outcomes of decision $\boldsymbol{a}_t$ for $\alpha \in (0,1)$ as follows:
\begin{definition} \textbf{(CVaR $\Phi_{\alpha} (\boldsymbol{a}_t)$):}
	\label{def:cvar}
Considering $\Phi_{\alpha}(\boldsymbol{a}_t) $ is the conditional expectation with the energy shortfall associated with the variable $\boldsymbol{a}_t$, and the energy shortfall is at least $\xi_{\alpha} (\boldsymbol{a}_t)$. Thus, the CVaR $\Phi_{\alpha} (\boldsymbol{a}_t)$ is defined as follows: 
\begin{equation} \label{eq:cvar_eq}
	 \Phi_{\alpha}(\boldsymbol{a}_t) = \underset{\xi \in \mathbb{R}} \min \; \frac{1}{(1-\alpha)} \mathbb{E}_{\xi_{\alpha} (\boldsymbol{a}_t)}\big[\Upsilon(\boldsymbol{a}_t,\boldsymbol{s}_t) \big],
	 \end{equation}
where the probability of energy shortfall $P(\Upsilon(\boldsymbol{a}_t,\boldsymbol{s}_t)) \ge P(\xi_{\alpha} (\boldsymbol{a}_t))$ is equal to $1-\alpha$.
\end{definition} 
To characterize both VaR $\xi_{\alpha} (\boldsymbol{a}_t)$ (in Definition \ref{def:var}) and CVaR  $\Phi_{\alpha}(\boldsymbol{a}_t)$ (in Definition \ref{def:cvar}) in terms of a function $H_{\alpha}(\boldsymbol{a}_t, \xi)$ on $ \boldsymbol{A}_t \times \mathbb{R}$, we define as follows:
\begin{equation} \label{eq:var_cvar_eq}
H_{\alpha}(\boldsymbol{a}_t, \xi) = \underset{\xi \in \mathbb{R}} \min \; \xi + \frac{1}{(1-\alpha)} \mathbb{E}_{\xi \sim \psi(\boldsymbol{a}_t, \xi)} \big[\Upsilon(\boldsymbol{a}_t,\boldsymbol{s}_t) - \xi\big]^{+},
\end{equation}
where $\mathbb{E} [\Upsilon(\boldsymbol{a}_t,\boldsymbol{s}_t) - \xi]^{+}$ is positive and as a function of $\xi$, $H_{\alpha}(\boldsymbol{a}_t, \xi)$ is continuous and differentiable \cite{IEEEhowto:Rockafellar_cvar_base,IEEEhowto:Krokhmal_cvar_expline,IEEEhowto:Moazeni_CVar_Power_system}. Therefore, we represent the risk assessment of risk-aware energy scheduling problem as follows:
\begin{equation} \label{eq:cvar_eq_final}
\underset{\xi \in \mathbb{R}} \min \;  H_{\alpha}(\boldsymbol{a}_t, \xi).
\end{equation}
In later section, we will provide an optimization problem for the microgrid controller so that the energy scheduling of the MEC network can be calculated efficiently.

\subsection{Problem Formulation}
The goal of the proposed risk-aware energy scheduling problem is to reduce the difference between the estimated generation and demand of the microgrid-powered MEC network while satisfying the CVaR confidence level. In particular, the objective is to minimize the expected energy residual (i.e., loss) between estimated demand and supply by capturing the tail-risk of the uncertain energy generation and consumption. Thus, the risk-aware energy scheduling problem consists of two decision variables: 1) storage or buying decision $a_t$, and 2) probability bound of CVaR $\xi$. The problem is formulated as follows:
\begin{subequations}\label{Opt_1_1}
	\begin{align}
	\underset{a_t, \xi } \min
	&\; \sum_{t \in \mathcal{T}} \sum_{i \in \mathcal{B}} \sum_{j \in \mathcal{C}_i} \sum_{k \in \mathcal{K}} \Omega_{ik}(t)\Upsilon(\boldsymbol{a}_t,\boldsymbol{s}_t)  \tag{\ref{Opt_1_1}}, \\
	\text{s.t.} \quad & \label{Opt_1_1:const1} P(H_{\alpha}(\boldsymbol{a}_t, \xi)) \le \omega,\forall t \in \mathcal{T}, \\
	&\label{Opt_1_1:const2} (a_t -1) \mathcal{E}^{\textrm{dem}}(t) \le a_t\mathcal{E}^{\textrm{gen}}(t), \forall i \in \mathcal{B}, \forall j \in \mathcal{C}_i, t \in \mathcal{T},\\
	\begin{split}
	&\label{Opt_1_1:const3} a_t {\mathcal{E}}^{\textrm{sto}}(t) + (1-a_t){\mathcal{E}}^{\textrm{buy}}(t) \ge 0, \forall i \in \mathcal{B},t \in \mathcal{T},
	\end{split}\\
	&\label{Opt_1_1:const4} (a_t -1) {\mathcal{E}}^{\textrm{non}}(t) \le    a_t {\mathcal{E}}^{\textrm{sto}}(t),\forall i \in \mathcal{B},t \in \mathcal{T},\\
	&\label{Opt_1_1:const5} \sum_{t \in \mathcal{T}} a_t \le T,  \\
	&\label{Opt_1_1:const6}a_t \in \left\lbrace0,1 \right\rbrace, \forall t \in \mathcal{T}, \\ 
	&\label{Opt_1_1:const7} {a_t\mathcal{E}}^{\textrm{sto}}(t) \le {\mathcal{E}}^{\textrm{sto}}_{\textrm{max}}(t) \le {\mathcal{E}}^{\textrm{gen}}_{\textrm{max}}(t) , \forall t \in \mathcal{T}.
	\end{align}
\end{subequations}
In problem \eqref{Opt_1_1}, the objective function contains multiplication of the task association indicator $\Omega_{ik}(t)$ and loss function \eqref{eq:loss_fn}, where $\Omega_{ik}(t)$ is given for task $k$ and is associated with SBS $i$. Constraint \eqref{Opt_1_1:const1} states inequalities for any $\xi \in \mathbb{R} \colon P(\Phi_{\alpha}(\boldsymbol{a}_t)) \le P(H_{\alpha}(\boldsymbol{a}_t, \xi)) \le \omega$, where $\omega$ denotes the risk tolerance and takes a small value. As a result, the probability of energy scheduling loss satisfies the maximum tolerable risk $P(\Upsilon(\boldsymbol{a}_t,\boldsymbol{s}_t)) \le \omega$. Therefore, maintaining the relationship between energy demand and supply is very crucial due to the nondeterministic nature of both energy consumption and generation. Hence, constraint \eqref{Opt_1_1:const2} ensures coupling between the energy demand $\mathcal{E}^{\textrm{dem}}(t)$ and generation $\mathcal{E}^{\textrm{gen}}(t)$, where the energy demand is derived from the univariate (single-variable) quadratic function \eqref{eq:All_BS_total_eng} considering both network energy consumption \eqref{eq:comm_energy} and MEC server computational energy consumption \eqref{eq:cpu_energy}. The amount of energy storage ${\mathcal{E}}^{\textrm{sto}}(t)$ and buying ${\mathcal{E}}^{\textrm{buy}}(t)$ depends on a decision of the binary decision variable $a_t$, which is defined in \eqref{eq:zeta_decision_variable}. Therefore, constraint \eqref{Opt_1_1:const3} takes into account the renewable energy generation ${\mathcal{E}}^{\textrm{ren}}(t)$ on the microgrid side for deciding energy buying and storing. The amount of storage energy ${\mathcal{E}}^{\textrm{sto}}(t)$ is calculated based on the availability of non-renewable energy generation ${\mathcal{E}}^{\textrm{non}}(t)$ at time slot $t$, where constraint \eqref{Opt_1_1:const4} ensures a coupling relation between the non-renewable energy generation ${\mathcal{E}}^{\textrm{non}}(t)$ and stored energy ${\mathcal{E}}^{\textrm{non}}(t)$. As a result, the decision variable $a_t$ completely relies on constraints \eqref{Opt_1_1:const3} and \eqref{Opt_1_1:const4} for determining the energy storage and buying decision. In formulated problem \eqref{Opt_1_1}, the decision variable $a_t$ is a binary variable, where constraint \eqref{Opt_1_1:const5} guarantees the energy store/buy decision for the entire time horizon $\mathcal{T}$ and constraint \eqref{Opt_1_1:const6} assures that the decision variable $a_t$ is a binary variable. Finally, constraint \eqref{Opt_1_1:const7} ensures that the total storage energy does not exceed the limit of maximum capacity.

Since formulated problem \eqref{Opt_1_1} is a mixed-integer programming problem with the corresponding constraints \eqref{Opt_1_1:const1} through \eqref{Opt_1_1:const6}, this problem can be reduced to a 0/1 multiple-knapsack problem as a base problem \cite{IEEEhowto:Knapsack_Lagoudakis}, which is NP-Complete \cite{IEEEhowto:Aaronson_NP_Complete}. Similar to the 0/1 multiple-knapsack problem, problem \eqref{Opt_1_1} is combinatorial in nature, which can be used to determine the feasible energy scheduling of the network by the microgrid; however, the complexity of problem \eqref{Opt_1_1} leads to an exponential complexity $O(2^{T \times B \times C \times K})$. In fact, constraint \eqref{Opt_1_1:const1} exhibits stochastic properties due to uncertainties in energy consumption and generation. Note that there is no known polynomial algorithm that can solve problem \eqref{Opt_1_1} with the optimal results. As a result, we can infer that problem \eqref{Opt_1_1} is NP-hard, similar to  the multiple-knapsack problems \cite{ IEEEhowto:ang_NP_complete}. 

To obtain a solution of problem \eqref{Opt_1_1}, we model an $N$-agent stochastic game with multi-agent deep reinforcement learning approaches. This model consists of two parts: 1) global agent (i.e., critic), and 2) virtual agents (i.e., actors). The global agent works for exploration which guides the virtual agents to correct the policy estimation by aggregating the each virtual agent's outcome (i.e., observation including current state, reward, action, and next state). Thus, virtual agents learn environment (i.e., $\boldsymbol{s}_t$) in parallel (asynchronous way) and explore with the suggestions (i.e., temporal difference among the value functions) of global agent. Each virtual agent interdependently explores and predicts the next state (i.e., $\boldsymbol{s}_{t+1}$) by following the exploration suggestions from the global agent. This mechanism amplifies the learning process with more diverse information. As a result, global agent is capable of choosing the best energy schedule policy by employing a joint decision for determining the decision $a_t$ of the problem \eqref{Opt_1_1}. In this solution, we employ shared neural networks (weight sharing) for low computational complexity among the virtual agent and global agent. The MADRL model accepts the best policy among the other agents toward updating the global energy schedule policy with less information. A detailed discussion of the solution to the risk-aware energy scheduling problem is given in the following section.

\begin{figure}[!t]
	\centerline{\includegraphics[scale=.5]{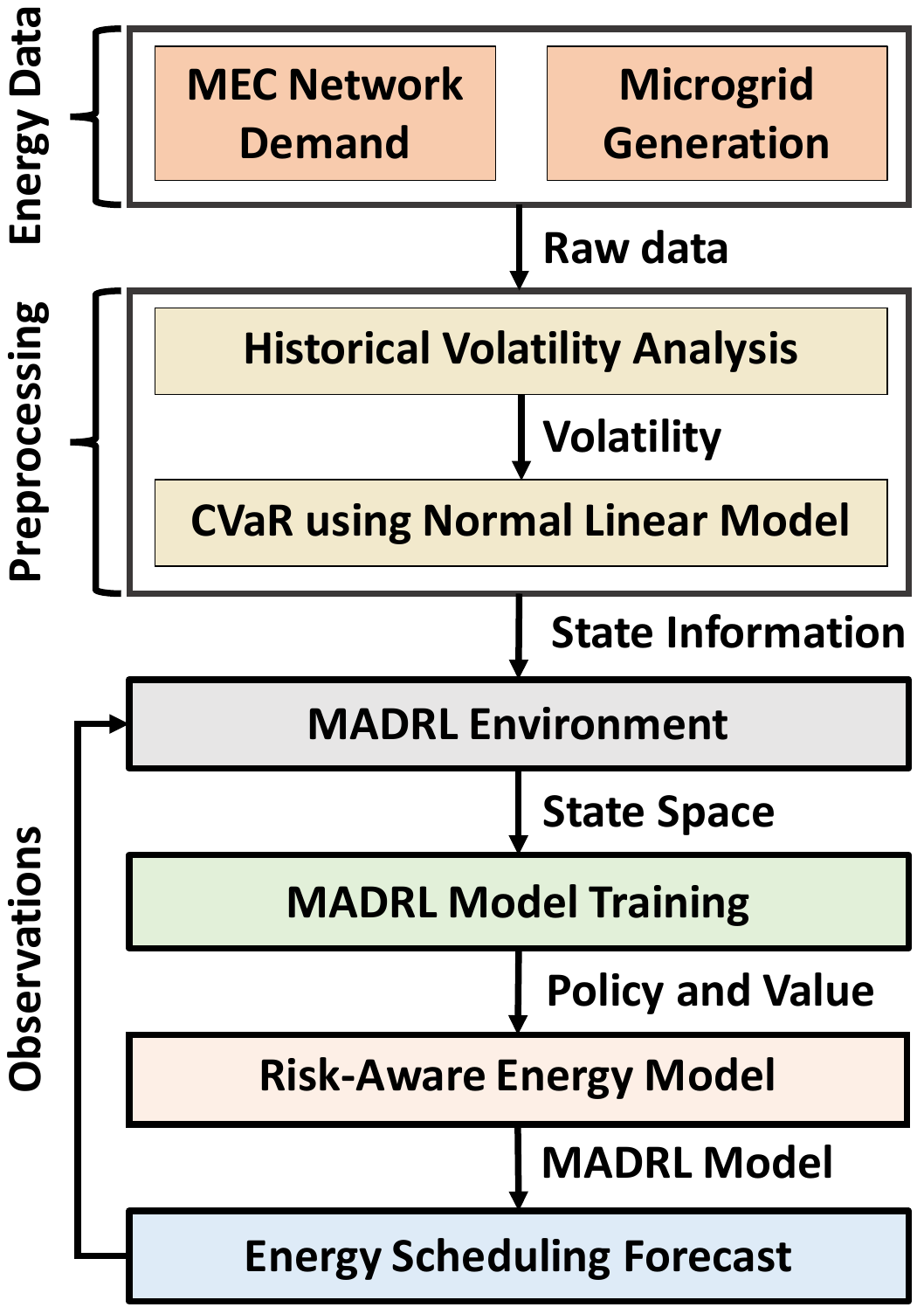}}
	\caption{Risk-aware energy scheduling solution procedure.}
	\label{fig:sol}
\end{figure}  
\section{Risk-Aware Energy Scheduling via an $N$-Agent Stochastic Game and MADRL}
In this section, first we devise an $N$-agent discounted rewards stochastic game model for the risk-aware energy scheduling problem in \eqref{Opt_1_1}, where we show that this game has at least one Nash equilibrium thus ensuring an optimal energy scheduling policy. Second, we solve this game using the MADRL-based Asynchronous A3C approach. The overall solution approach is shown in Fig. \ref{fig:sol} and a detailed discussion of the $N$-agent stochastic game and MADRL-based risk-aware energy scheduling are explained later in this section. 

\subsection{$N$-Agent Stochastic Game with MADRL for Risk-Aware Energy Scheduling}
First, we convert the objective of risk-aware energy scheduling problem \eqref{Opt_1_1} from a loss minimization problem to a reward maximization problem. To do this, we consider a multi-agent reinforcement learning setting with a set of virtual agents $\mathcal{N} = \left\{{1,2,\dots,N}\right\}$ defined by a set of states $\mathcal{S} = \left\{{1,2,\dots,S}\right\}$, a set of agent actions $\mathcal{A} = \left\{{\mathcal{A}_1,\mathcal{A}_2,\dots,\mathcal{A}_N}\right\}$, and a set of observations $\mathcal{O} = \left\{{\mathcal{O}_1,\mathcal{O}_2,\dots,\mathcal{O}_N}\right\}$ for each agent \cite{IEEEhowto:Littman_Markov_games}. From now on in the entire paper, virtual agents are represented as agents. The state-space at time slot $t$ is redefined by a tuple $\boldsymbol{s}_t \coloneqq (\mathcal{E}^{\textrm{dem}}(t), \mathcal{E}^{\textrm{ren}}(t), \mathcal{E}^{\textrm{sto}}(t), P(H_{\alpha}(\boldsymbol{a}_t, \xi))) \in \mathcal{S}$, where $\mathcal{E}^{\textrm{dem}}(t), \mathcal{E}^{\textrm{ren}}(t), \mathcal{E}^{\textrm{sto}}(t)$, and $P(H_{\alpha}(\boldsymbol{a}_t, \xi))$ are the energy demand, renewable energy, stored energy, and probability of CVaR, respectively. To calculate $P(H_{\alpha}(\boldsymbol{a}_t, \xi))$, we use the normal linear model as shown in Algorithm \ref{alg:CvaR}. In Algorithm \ref{alg:CvaR}, first, we calculate VaR (lines 3 and 4), while satisfying $P(\xi_{\alpha} (\boldsymbol{a}_t)) \ge 1- \alpha$ (constraint \eqref{Opt_2_1:const1}). Second, using the distribution of the calculated VaR, we determine the probability of CVaR $P(H_{\alpha}(\boldsymbol{a}_t, \xi))$ (lines 5 and 6 in Algorithm \ref{alg:CvaR}) and update the state-space information in line 8. Finally, this algorithm returns the updated state-space (in line 10) for further use. We redefine vector $\boldsymbol{a}_t$ as an action space $\boldsymbol{a}_t \in \mathcal{A}$ that is comprised of ${(\zeta_t^1, \zeta_t^0)}$. Where $ \zeta_t^1$ and $\zeta_t^0$ are discrete variables such that $\zeta_t^1$ determines the action regarding energy storage for the microgrid and $\zeta_t^0$ decides the action for energy buying from the main grid at time slot $t$.
\begin{algorithm}[t!]
	\caption{CVaR Calculation Using Normal Linear Model}
	\label{alg:CvaR}
	\begin{algorithmic}[1]
		\renewcommand{\algorithmicrequire}{\textbf{Input:}}
		\renewcommand{\algorithmicensure}{\textbf{Output:}}
		\REQUIRE  $ \mathcal{E}^{\textrm{dem}}(t), \mathcal{E}^{\textrm{ren}}(t), \mathcal{E}^{\textrm{non}}(t), \mathcal{E}^{\textrm{sto}}(t)$, $T$, $\alpha$, $\xi$
		\ENSURE  $\forall \boldsymbol{s}_t \in \mathcal{S} $ 
		\\ \textbf{Initialization}: \textit{mu, sig, volatility }
		\FOR {$\forall t \in \mathcal{T}$}
		\FOR {\textbf{Until:} $P(\xi_{\alpha} (\boldsymbol{a}_t)) \ge 1- \alpha$, Constraints: \eqref{Opt_2_1:const1}}  
		\STATE \textit{sig = $volatility * \sqrt{\frac{1}{T}}$}
		\STATE \textit{VaR} = $norm.ppf(1-\alpha) * sig - mu$
		\STATE \textit{CVaR} = $\frac{1}{(1-\alpha)} * norm.pdf(VaR) * sig - mu$
		\STATE \textbf{Assign:} $P(H_{\alpha}(\boldsymbol{a}_t, \xi)) = P(\textit{CVaR})$
		\ENDFOR
		\STATE \textbf{Update:} $\boldsymbol{s}_t \coloneqq ( \mathcal{E}^{\textrm{dem}}(t), \mathcal{E}^{\textrm{ren}}(t), \mathcal{E}^{\textrm{sto}}(t), P(H_{\alpha}(\boldsymbol{a}_t, \xi)) )$  
		\ENDFOR
		\RETURN $\forall \boldsymbol{s}_t \in \mathcal{S} $, $\forall t \in \mathcal{T}$ 
	\end{algorithmic} 
\end{algorithm}

The energy store/buying action for each agent $n \in \mathcal{N}$ with parameter $\theta_{n}$ is determined by a stochastic policy $\pi_{\theta_{n}}$, where $\pi_{\theta_{n}} \colon \mathcal{O}_n \times \mathcal{A}_n \mapsto [0,1]$. A state transition function\footnote{For a state space $\boldsymbol{s}_t \in \mathcal{S}$, $\mathcal{A}_1 \times \mathcal{A}_2, \dots \times \mathcal{A}_N$ is the joint action space, which is the Cartesian product of the $N$ virtual action spaces $\mathcal{A}$. Thus, the transition probability $\Gamma$ for next state $\boldsymbol{s}_{t+1} \in \mathcal{S}$ is determined by $\Gamma = \prod_{n \in \mathcal{N}} \Gamma_n(\boldsymbol{s}_{t} \in \mathcal{S},{A}_n, \boldsymbol{s}_{t+1} \in \mathcal{S})$ \cite{IEEEhowto:Lowe_Multi_agent_actor_critic, IEEEhowto:Littman_Markov_games} due to each virtual agent transition is independent.} $\Gamma \colon \mathcal{S} \times \mathcal{A}_1 \times \mathcal{A}_2, \dots \times \mathcal{A}_N \mapsto \mathcal{S}$ determines the next state according to policy $\pi_{\theta_{n}}$. Each agent $n$ determines the reward as a function of the state and its action $r_n \colon \mathcal{S} \times \mathcal{A}_n \mapsto \mathbb{R}$, which is an element of the observation tuple $\boldsymbol{o}_{n}$ for agent $n$, where $\boldsymbol{o}_{n} \coloneq (\boldsymbol{s}_{t}, \boldsymbol{a}_{t}, r_{t}, \boldsymbol{s}_{t'}) $. This observation correlates with the state space $\mathcal{S}$ such that $\boldsymbol{o}_{n} \colon \mathcal{S} \mapsto \mathcal{O}_n$. The agent follows a discrete time slot $t \in\left\{1,2, \dots, T \right\}$, and the objective of the each agent $n$ is to maximize the total expected reward defined by \cite{IEEEhowto:Lowe_Multi_agent_actor_critic}, 
\begin{equation} \label{eq:each_reward}
r_{n}(\boldsymbol{a}_{t}, \boldsymbol{s}_{t}) = \underset{\boldsymbol{a}_{t} \in \mathcal{A}_n} \max \; \mathbb{E}\big[\sum_{t'=t}^{\infty} \gamma^{t'-t} r_{t}(\boldsymbol{a}_{t}, \boldsymbol{s}_{t})\big],
\end{equation}
where $\gamma \in (0,1)$ is a discount factor and each reward $r_{t}(\boldsymbol{a}_{t}, \boldsymbol{s}_{t}) = a_t$ using \eqref{eq:zeta_decision_variable} (i.e., if the microgrid is able to fulfill energy demand from its own sources, and $0$ otherwise).

The expectation of the action value function for agent $n$ taking action $\boldsymbol{a}_t$ in state $\boldsymbol{s}_t$ is defined as follows:
\begin{equation} \label{eq:each_action_value}
	Q^{\pi_{\theta_{n}}}(\boldsymbol{s}_{t},\boldsymbol{a}_{t}) = \mathbb{E}_{\pi_{\theta_{n}}}\big[\sum_{t'=t}^{\infty} \gamma^{t'-t} r_{t}(\boldsymbol{a}_{t}, \boldsymbol{s}_{t}) |\boldsymbol{s}_{t}, \boldsymbol{a}_{t}\big],
\end{equation}
where $\gamma^{t'-t}$ ensures convergence of the action value function $Q^{\pi_{\theta_{n}}}(\boldsymbol{s}_{t},\boldsymbol{a}_{t})$ estimation by controlling the discount factor over the infinity time horizon. Thus, the expectation of the state value function for agent $n$ is given below:
\begin{equation} \label{eq:each_state_value}
	V^{\pi_{\theta_{n}}}(\boldsymbol{s}_{t}) = \mathbb{E}_{\boldsymbol{a}_t \sim \pi_{\theta_{n}}(\boldsymbol{a}_{t}| \boldsymbol{s}_{t})}\big[Q^{\pi_{\theta_{n}}}(\boldsymbol{s}_{t},\boldsymbol{a}_{t})\big],
\end{equation}
where the action value function $V^{\pi_{\theta_{n}}}(\boldsymbol{s}_{t})$ is determined by \eqref{eq:each_action_value}. In this case, the environment of the RL problem is non-trivial and in general, the transition probability between the two consecutive states $\boldsymbol{s}_{t}$ and $\boldsymbol{s}_{t+1}$ (for simplicity, we change the notation of next state from $\boldsymbol{s}_{t+1}$ to $\boldsymbol{s}_{t'}$) with some action is unknown. The model-free reinforcement learning approach is appropriate to learn the dynamics of the problem environment \cite{IEEEhowto:Mnih_async_a3c}. To solve the RL problem, the goal is to find the optimal policy $\pi_{\theta_{n}}^*(\boldsymbol{a}_{t}|\boldsymbol{s}_{t})$ for value function \eqref{eq:each_state_value}, where the optimal state value is as follows:
\begin{equation} \label{eq:opt_val_action_fn}
\begin{split}
V^{\pi_{\theta_{n}}}(\boldsymbol{s}_{t})  = \;\;\;\;\;\;\;\;\;\;\;\;\;\;\;\;\;\;\;\;\;\;\;\;\;\;\;\;\;\;\;\;\;\;\;\;\;\;\;\;\;\;\;\;\;\;\;\;\;\;\;\;\;\;\;\;\;\;\;\; \\ \underset{\boldsymbol{a}_{t} \in \mathcal{A}} \max \;\mathbb{E}_{\pi_{\theta_{n}}^*} \big[\sum_{n \in N}r_{n}(\boldsymbol{a}_{t'}, \boldsymbol{s}_{t'}) + \sum_{t'=t}^{\infty} \gamma^{t'-t} V_{t'}^{\pi_{\theta_{n}}}(\boldsymbol{s}_{t'}) |\boldsymbol{s}_{t}, \boldsymbol{a}_{t}\big].
\end{split}
\end{equation}
Consequently, problem \eqref{Opt_1_1} can be rewritten as follows: 
\begin{subequations}\label{Opt_2_1}
	\begin{align}
	\underset{\boldsymbol{a}_t \in \mathcal{A}, \pi_{\theta_{n}}} \max
	&\; \sum_{t \in \mathcal{T}} \sum_{i \in \mathcal{B}} \sum_{j \in \mathcal{C}_i} \sum_{k \in \mathcal{K}} \Omega_{ik}(t)V^{\pi_{\theta_{n}}}(\boldsymbol{s}_{t})   \tag{\ref{Opt_2_1}}, \\
	\text{s.t.} \quad & \label{Opt_2_1:const1} P(\Upsilon(\boldsymbol{a}_t,\boldsymbol{s}_t)) \ge (1-\alpha), \\
	 &\label{Opt_2_1:const2} \eqref{Opt_1_1:const2} \; \text{to} \; \eqref{Opt_1_1:const7}. 
	\end{align}
\end{subequations}
In problem \eqref{Opt_2_1}, we admit a new constraint \eqref{Opt_2_1:const1}, for a given probability $\alpha$, state $\boldsymbol{s}_t$ and action $\boldsymbol{a}_t$ this constraint satisfies the CVaR confidence level of loss function \footnote{In order to formulate the reinforcement learning in \eqref{Opt_2_1}, we additionally include the probability of energy shortfall $P(\Upsilon(\boldsymbol{a}_t,\boldsymbol{s}_t))$ as a new constraint \eqref{Opt_2_1:const1} with a given CVaR confidence value of $1-\alpha$. We solve \eqref{Opt_2_1:const1} by Algorithm \ref{alg:CvaR} and append the CVaR risk $P(H_{\alpha}(\boldsymbol{a}_t, \xi)$ with the state space $\boldsymbol{s}_t \coloneqq ( \mathcal{E}^{\textrm{dem}}(t), \mathcal{E}^{\textrm{ren}}(t), \mathcal{E}^{\textrm{sto}}(t), P(H_{\alpha}(\boldsymbol{a}_t, \xi)))$. Therefore, the risk of the energy shortfall for the loss \eqref{eq:loss_fn} is considered as an \emph{input} to the \emph{reward/value} function. Thus, the proposed model finds a stochastic policy $\pi_{\theta_{n}}$ that is mapped to the observation $\pi_{\theta_{n}} \colon \mathcal{O}_n \times \mathcal{A}_n \mapsto [0,1]$, where $\boldsymbol{o}_{n} \colon (\boldsymbol{s}_{t}, \boldsymbol{a}_{t}, r_{t}, \boldsymbol{s}_{t'}) \in \mathcal{O}_n$.} \eqref{eq:loss_fn}. Further, $\boldsymbol{a}_t \in \mathcal{A}$ and $\pi_{\theta_{n}}$ are two decision variables, where $\boldsymbol{a}_t$ decides energy storing or buying decision and $\pi_{\theta_{n}}$ represents the energy scheduling policy with parameters $\theta$. The constraints from \eqref{Opt_1_1:const2} to \eqref{Opt_1_1:const7} remain the same as in problem \eqref{Opt_1_1}. All though the objective has changed in problem \eqref{Opt_2_1}, the complexity is the same as the base problem \eqref{Opt_1_1}. Therefore, we discretize the risk-aware energy scheduling problem \eqref{Opt_2_1} using the $N$-agent stochastic game. The definition of the game is as follows:
\begin{definition} \textbf{($N$-agent Stochastic Game):}
\label{def:game}
An $N$-agent (player) stochastic game $\mathcal{G}$ consists of a tuple ${( \mathcal{S}, \mathcal{A}_1,\mathcal{A}_2,\dots,\mathcal{A}_N, r_{1},r_{2}, \dots, r_{N}, \Gamma )}$, where $\mathcal{S}$ is the state space, $\mathcal{A}_n$ determines the action space of agent $n \in \mathcal{N}$, the reward (payoff) is defined by $r_{n} \colon \mathcal{S} \times \mathcal{A}_1 \times \mathcal{A}_2, \dots \times \mathcal{A}_N \mapsto \mathbb{R}$, and $\Gamma \mapsto [0,1]$ is the state transition probability of agent $n$ at state space $\mathcal{S}$. 
\end{definition}
For a state $\boldsymbol{s}_{t} \in \mathcal{S}$ at time $t$, all agents $n \in \mathcal{N}$ independently choose their own actions $(\boldsymbol{a}_{t1}, \boldsymbol{a}_{t2} , \dots,\boldsymbol{a}_{tN})$ and determine the rewards $(r_{1}(\boldsymbol{a}_{t}, \boldsymbol{s}_{t}), r_{2}(\boldsymbol{a}_{t}, \boldsymbol{s}_{t}), \dots,r_{N}(\boldsymbol{a}_{t}, \boldsymbol{s}_{t}))$. After that, for a fixed transition probability, the current state $\boldsymbol{s}_{t} \in \mathcal{S}$ transit to the next state $\boldsymbol{s}_{t'} \in \mathcal{S}$ and satisfies the following property:
\begin{equation} \label{eq:game_transition_prob}
\begin{split}
\sum_{\boldsymbol{s}_{t'} \in \mathcal{S}} P(\boldsymbol{s}_{t'}|\boldsymbol{s}_{t}, \boldsymbol{a}_{t1}, \boldsymbol{a}_{t2} , \dots,\boldsymbol{a}_{tN}) = 1.
\end{split}
\end{equation}
In this game, we determine a policy $\pi_{\theta_{n}}$ by updating the parameters $\theta_{n}$, where for time slot $t$, parameter $\theta_{t}$ is derived via an $N$-agent stochastic game $\mathcal{G}$, and action value function \eqref{eq:each_action_value} can be represented as follows:  
\begin{equation} \label{eq:opt_state_action_param_fn}
Q^{\pi_{\theta_{n}}}(\boldsymbol{s}_{t},\boldsymbol{a}_{t})  \approx Q^{\pi_{\theta_{n}}}(\boldsymbol{s}_{t},\boldsymbol{a}_{t}; \theta_{t}).
\end{equation}
Hence, state value function \eqref{eq:opt_val_action_fn} can be rewritten as follows:
\begin{equation} \label{eq:opt_val_action_param_fn}
V^{\pi_{\theta_{n}}^*}(\boldsymbol{s}_{t})  \approx V^{\pi_{\theta_{n}}}(\boldsymbol{s}_{t};\theta_{t}).
\end{equation}
Moreover, the parameterized policy is represented as follows:
\begin{equation} \label{eq:param_policy_fn}
\pi_{\theta_{n}}(\boldsymbol{a}_{t}|\boldsymbol{s}_{t}) \approx \pi_{\theta_{n}}(\boldsymbol{a}_{t}|\boldsymbol{s}_{t};\theta_{t}).
\end{equation}
Thus, if $\pi_{\theta_{n}}$ is the parameterized policy for agent $n$ at current state $\boldsymbol{s}_{t}$ then the value function can be redefined as follows:
\begin{equation} \label{eq:game_dis_count_state_value}
\begin{split}
V^{\pi_{\theta_{n}}}(\boldsymbol{s}_{t}, \pi_{\theta_{1}},\pi_{\theta_{2}},\dots, \pi_{\theta_{N}} ) = \underset{\boldsymbol{a}_{t} \in \mathcal{A}} \max \; \sum_{t'=t}^{\infty} \gamma^{t'-t} \mathbb{E}_{\pi_{\theta_{n}}} \big[r_{n}(\boldsymbol{a}_{t}, \boldsymbol{s}_{t})|\mathcal{O}_n\big],
\end{split}
\end{equation}
where $\mathcal{O}_n = (\boldsymbol{s}_{t}, \pi_{\theta_{1}},\pi_{\theta_{2}},\dots, \pi_{\theta_{N}} ), \forall n \in \mathcal{N}$ and policy definition is given as follows:   
\begin{definition}\textbf{(Game Policy):}
	\label{def:game_strategy}
	The policy (strategy) $\pi_{\theta_{n}}$ of the $N$-agent stochastic game is defined by $\pi_{\theta_{n}} \colon \mathcal{O}_n \times \mathcal{A}_n \mapsto [0,1]$. Here, we consider a stationary policy such that state $\boldsymbol{s}_{t} \in \mathcal{S}$ remains unchanged during time $t$ and the policy is determined by $\pi_{\theta} = (\pi_{\theta_{1}},\pi_{\theta_{2}},\dots, \pi_{\theta_{N}})$.    
\end{definition}
In the $N$-agent stochastic game $\mathcal{G}$, a Nash equilibrium determines a joint strategy, where each agent achieves the best response to the other agents. The policy (strategy) of each agent is defined over time $t \in \mathcal{T}$ and the definition of policy Nash equilibrium is as follows: 
\begin{definition}
	\label{def:game_equilibrium_strategy} \textbf{(Policy Nash Equilibrium):}
	In game $\mathcal{G}$, for the state space $\boldsymbol{s}_{t} \in \mathcal{S}$, a Nash equilibrium contains a tuple of $\forall n \in \mathcal{N}$ policies $(\pi_{\theta_{1}^*},\pi_{\theta_{2}^*},\dots, \pi_{\theta_{N}^*})$, and for agent $n$,
	\begin{equation} \label{eq:game_equilibrium_strategy_const}
	\begin{split}
	V^{\pi_{\theta_{n}}}(\boldsymbol{s}_{t}, \pi_{\theta_{1}}^*,\pi_{\theta_{2}}^*,\dots, \pi_{\theta_{N}}^* ) \ge \;\;\;\;\;\;\;\;\;\;\; \;\;\;\;\;\;\;\;\;\;\\ V^{\pi_{\theta_{n}}} (\boldsymbol{s}_{t}, \pi_{\theta_{1}}^*,\dots, \pi_{\theta_{n-1}}^*, \pi_{\theta_{n}}, \pi_{\theta_{n+1}}^*, \dots, \pi_{\theta_{N}}^*).
	\end{split}
	\end{equation}     
\end{definition}
The \eqref{eq:opt_val_action_fn} determines optimal state value function, which is equivalent to the Nash equilibrium for an agent $n$ with the Nash equilibrium policy $\pi_{\theta_{n}}^*$. A joint policy Nash equilibrium of the $N$-agent stochastic game $\mathcal{G}$ is as follows:
\begin{definition}
	\label{def:game_Q_function} \textbf{(Joint Policy Nash Equilibrium):}
	For a joint policy Nash equilibrium of agent $n$, the state-action value function is defined over $(\boldsymbol{s}_{t}, \boldsymbol{a}_{t1}, \boldsymbol{a}_{t2}, \dots,\boldsymbol{a}_{tN})$, where the state-action value function is the sum of agent $n$'s current reward addition with future rewards, and is defined as follows:  
	\begin{equation} \label{eq:game_Q_function}
	\begin{split}
	Q^{\pi_{\theta_{n}}^*}(\boldsymbol{s}_{t}, \boldsymbol{a}_{t1} , \dots,\boldsymbol{a}_{tN})  = 
	r_{n}(\boldsymbol{s}_{t}, \boldsymbol{a}_{t1} , \dots,\boldsymbol{a}_{tN}) \; + \;\;\;\;\;\;\;\;\;\;\;\;\;\;\;\; \\ \sum_{\boldsymbol{s}_{t'} \in \mathcal{S}, t'=t}^{\infty} \gamma^{t'-t} P(\boldsymbol{s}_{t'}|\boldsymbol{s}_{t}, \boldsymbol{a}_{t1}, \dots,\boldsymbol{a}_{tN}) V^{\pi_{\theta_{n}}}(\boldsymbol{s}_{t'}, \pi_{\theta_{1}}^*,\dots, \pi_{\theta_{N}}^* ). 
\end{split}
\end{equation}    
\end{definition}  
In \eqref{eq:game_Q_function}, $(\pi_{\theta_{1}}^*,\dots, \pi_{\theta_{N}}^*)$ determines the joint policy Nash equilibrium. The current reward for agent $n$ is defined by $r_{n}(\boldsymbol{s}_{t}, \boldsymbol{a}_{t1}, \dots,\boldsymbol{a}_{tN})$, and the total discounted reward for state space $\boldsymbol{s}_{t'} \in \mathcal{S}$ with the joint action $(\boldsymbol{a}_{t1} , \dots,\boldsymbol{a}_{tN})$ is represented by $V^{\pi_{\theta_{n}}}(\boldsymbol{s}_{t'}, \pi_{\theta_{1}}^*,\dots, \pi_{\theta_{N}}^* )$. As a result, \eqref{eq:game_Q_function} follows a joint policies (strategies) Nash equilibrium, in which the optimal state-action value (payoff) is determined by all the actions $(\boldsymbol{a}_{t1}, \dots,\boldsymbol{a}_{tN})$ from other agents in the state space  $\boldsymbol{s}_{t} \in \mathcal{S}$. In the multi-agent RL environments, agent $n$ not only observes its own reward, however also needs to know about the others observations as well. For state-action value function $Q^{\pi_{\theta_{n}}^*}(\boldsymbol{s}_{t}, \boldsymbol{a}_{t1} , \dots,\boldsymbol{a}_{tN})$, the joint policy is determined by $(\boldsymbol{a}_{t1} , \dots,\boldsymbol{a}_{tN})\sim \pi_{\theta}((\boldsymbol{a}_{t1} , \dots,\boldsymbol{a}_{tN}) | \boldsymbol{s}_{t}) = \underset{\boldsymbol{a}_{tn}} \argmax \; Q^{\pi_{\theta_{n}}^*}(\boldsymbol{s}_{t}, \boldsymbol{a}_{t1} , \dots,\boldsymbol{a}_{tN})$, where the chosen action represents the action of the original problem \eqref{Opt_1_1}. Thus, the game $\mathcal{G}$ can be decomposed by $N$-agent stage game and the definition is as follows:
\begin{definition} \textbf{($N$-agent Stage Game):}
	\label{def:game_stage_game}
	The $N$-agent (player) stochastic game $\mathcal{G}$ is represented with an $N$-agent stage game $({M}_1(\mathcal{O}_1, \boldsymbol{a}_{t}),{M}_2(\mathcal{O}_2, \boldsymbol{a}_{t}),\dots, {M}_N(\mathcal{O}_N, \boldsymbol{a}_{t}) )_t$, $\forall t \in \mathcal{T}$, under the state space  $\boldsymbol{s}_{t} \in \mathcal{S}$. For a single stage, the reward (payoff) of agent $n$ is determined by ${M}_n(\mathcal{O}_n, \boldsymbol{a}_{t})$ and the reward consists of the joint action and agent $n$'s own reward $r_{n}$. The reward of agent $n$ is defined as follows: 
	\begin{equation} \label{eq:n_agent_stage_game}
	\begin{split}
	{M}_n(\mathcal{O}_n, \boldsymbol{a}_{t}) = \;\;\;\;\;\;\;\;\;\;\;\;\;\;\;\;\;\;\;\;\;\;\;\;\;\;\;\;\;\;\;\;\;\;\;\;\;\;\;\;\;\;\;\;\;\;\;\;\;\;\;\;\;\;\;\;\;\;\;\;\;\; \\ \mathbb{E}_{\mathcal{O}_n,\boldsymbol{a}_{t} \sim \mathcal{M}_n} \big[r_{n}(\boldsymbol{a}_{t1}, \dots,\boldsymbol{a}_{tN})|(\boldsymbol{a}_{t1} \in \mathcal{A}_1, \dots,\boldsymbol{a}_{tN} \in \mathcal{A}_N)\big].
	\end{split}
	\end{equation} 
\end{definition}
We define $\vartheta_{\theta_{\bar{n}}}$ as the product of all agents policies (strategies) except for agent $n$, i.e., $\vartheta_{\theta_{\bar{n}}} = (\vartheta_{\theta_{1}} . \dots \vartheta_{\theta_{n-1}}.\vartheta_{\theta_{n+1}} . \dots\vartheta_{\theta_{N}})$. Hence, Nash equilibrium of $N$-agent stage game is defined as follows:
\begin{definition} \textbf{($N$-agent Stage Game Nash Equilibrium):}
	\label{def:game_joint_strategy_equilibrium}
	For the $N$-agent single stage game $({M}_1(\mathcal{O}_1, \boldsymbol{a}_{t}), {M}_2(\mathcal{O}_2, \boldsymbol{a}_{t}), \dots, {M}_N(\mathcal{O}_N, \boldsymbol{a}_{t}) )$, a Nash equilibrium is discretized by a joint strategy, $(\vartheta_{\theta_{1}}\dots \vartheta_{\theta_{N}})$, such that
	\begin{equation} \label{eq:game_joint_strategy_equilibrium_const}
	\begin{split}
	  \vartheta_{\theta_{n}} \vartheta_{\theta_{\bar{n}}} {M}_n(\mathcal{O}_n, \boldsymbol{a}_{t})\ge \hat{\vartheta}_{\theta_{n}} \vartheta_{\theta_{\bar{n}}} {M}_n(\mathcal{O}_n, \boldsymbol{a}_{t}), \forall \vartheta_{\theta_{n}} \in \hat{\vartheta}_{\theta_{n}} (\mathcal{A}_n)
	\end{split}
	\end{equation}     
\end{definition}
From Definitions \ref{def:game_equilibrium_strategy} and  \ref{def:game_joint_strategy_equilibrium}, we determine the Nash equilibrium with policy $\pi_{\theta_{n}}^*$ for agent $n$ of the discounted reward stochastic game, and the joint strategy Nash equilibrium $\vartheta_{\theta_{n}}$ of the stage game, respectively. We conclude with the following theorem:   
\begin{thm}
	\label{thm:game_Nash_equilibrium_point}
	Every $N$-agent (player) discounted stochastic game consists of at least one Nash equilibrium point in stationary strategies \cite{IEEEhowto: Fink_Equilibrium_Th1}.
\end{thm}
Using Theorem \ref{thm:game_Nash_equilibrium_point}, we can derive the following proposition:
\begin{proposition}
	\label{pro:discounted_reward_game_Nash_equilibrium_point}
	$\pi_{\theta_{n}}^*$ is the optimal policy, which is an equilibrium point with the equilibrium rewards (payoffs) $V^{\pi_{\theta_{n}}}(\pi_{\theta_{1}}^*,\dots, \pi_{\theta_{N}}^* )$ for the discounted reward stochastic game [see Appendix \ref{apd:discounted_reward_game_Nash_equilibrium_point_proof}].
\end{proposition}
Proposition \ref{pro:discounted_reward_game_Nash_equilibrium_point} not only confirms the Nash equilibrium of the $N$-agent (player) stochastic game $\mathcal{G}$, but also restrains strong evidence of the similar arguments from the previously studied $N$-player stochastic games \cite{IEEEhowto: Fink_Equilibrium_Th1,IEEEhowto:Herings_Equilibrium_Th1_proof}. 

\subsection{Solution with Asynchronous Advantage Actor-Critic (A3C)}
\begin{figure}[!t]
	\centerline{\includegraphics[scale=.3]{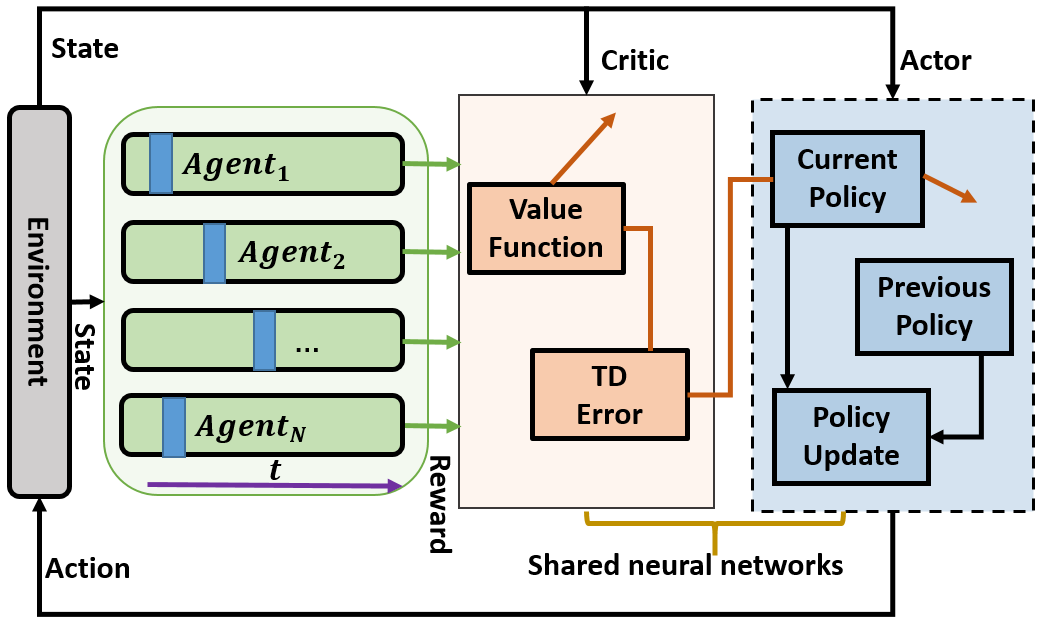}}
	\caption{Multi-agent deep reinforcement learning (MADRL) model for risk-aware energy scheduling \cite{IEEEhowto:Munir_GC_Multi_Agent}.}
	\label{fig:a3c}
	\vspace{-6mm}
\end{figure}
To solve problem \eqref{Opt_2_1} with fast and efficiently, we use the A3C model with shared neural networks (as seen in Fig. \ref{fig:a3c}). This approach determines the best policy estimation among the other agents, and is also capable of handling the curse of dimensionality of the state space, which is convenient for solving our formulated problem \eqref{Opt_2_1}.

In the A3C method, in order to learn the action value function $Q^{\pi_{\theta_{n}}^*}(\boldsymbol{s}_{t},\boldsymbol{a}_{t})$, and find the optimal policy $\pi_{\theta_{n}}^*$ using a deep Q-networks (DQN), the objective is to minimize the loss \cite{IEEEhowto:Mnih_deep_reinforcement_learning}:  
\begin{equation} \label{eq:loss_fn_dqn}
\mathcal{L}(\theta_{n}) =\mathbb{E}_{\boldsymbol{o}_n \in \mathcal{O}_n} \big[(Q^{\pi_{\theta_{n}}^*}(\boldsymbol{s}_{t},\boldsymbol{a}_{t}|\theta_{t}) - y_{t})^2\big],
\end{equation}
and the ideal target is represented as follows:
\begin{equation} \label{eq:ideal_target}
y_{t} = r_{n}(\boldsymbol{a}_{t}, \boldsymbol{s}_{t}) + \gamma \underset{a_{t'} \in \mathcal{A}  } \max \hat{Q}^{\pi_{\theta_{n}}^*}(\boldsymbol{s}_{t'},\boldsymbol{a}_{t'}),
\end{equation}
where  $\boldsymbol{o}_n\colon {( \boldsymbol{s}_{t}, \boldsymbol{a}_{t}, r_{t}, \boldsymbol{s}_{t'} )}$ is the observation with the current state $\boldsymbol{s}_{t'}$ and $\hat{Q}^{\pi_{\theta_{n}}^*}(\boldsymbol{s}_{t'},\boldsymbol{a}_{t'})$ represents the target value function. Both are important components for achieving a stable DQN learning process and parameters $\theta_{n}$ are periodically updated with the recent values.

In multi-agent reinforcement learning settings, policy $\pi_{\theta_{n}}$ is independently updated for each agent $n$ and this non-stationary nature violates the convergence characteristics of the learning process. The observations from the experience cannot be used for the general environment settings. To overcome these challenges, we use a policy gradient method to directly adjust parameters $\theta$ for policy $\pi_{\theta}$ to maximize the expected reward $\mathbb{E}[r_{n}]$. 

We consider a set of policies $\mathcal{\pi_{\theta}} = \left\{{\pi_{\theta_{1}},\pi_{\theta_{2}},\dots,\pi_{\theta_{N}}}\right\}$ with the parameter set $\mathcal{\theta} = \left\{{\theta_{1},\theta_{2},\dots,\theta_{N}}\right\}$ for $N$ agents and the expected return for policy $n$ is defined as follows:
\begin{equation} \label{eq:obj_policy_grdient}
J(\theta_{n}) = \underset{\pi_{\theta_{n}}} \max \; \mathbb{E}[r_{n}].
\end{equation}
The gradient of \eqref{eq:obj_policy_grdient} can be defined as follows:
\begin{equation} \label{eq:gradient_policy_grdient}
\nabla_{\theta_{n}} J(\theta_{n}) =  \mathbb{E}\big[\nabla_{\theta_{n}} \log \pi_{\theta_{n}}(\boldsymbol{a}_{n}|\boldsymbol{o}_{n}) Q_{n}^{\pi_{\theta}}(O, \boldsymbol{a}_{1}, \dots, \boldsymbol{a}_{N})\big],
\end{equation}
where $Q_{n}^{\pi_{\theta}}(\mathcal{O}, \boldsymbol{a}_{1}, \dots, \boldsymbol{a}_{N})$ is the centralized action-value function, $\boldsymbol{a}_{1}, \dots, \boldsymbol{a}_{N}$ determine all the actions for $N$ agents, and $\mathcal{O}$ represents all the observations $\mathcal{O} = \left\{ \boldsymbol{o}_{1}, \dots, \boldsymbol{o}_{N} \right\}$ for the $N$ agents. The gradient from equation \eqref{eq:gradient_policy_grdient} generates high bias and lower variance due to the deterministic observations. 

However, this model cannot be directly applied to this risk-aware energy scheduling scenarios because the risk of energy scheduling is highly dependant on uncertainties in both the energy consumption of the MEC networks and the renewable energy generation. To execute this energy scheduling model, we have approximated using $N$ continuous policies $\vartheta_{\theta_{n}}$ with respect to the parameters $\theta_{n}$, such that the policy gradient function can now be presented as follows:   
\begin{equation} \label{eq:new_gradient_policy_grdient}
\begin{split}
\nabla_{\theta_{n}} J(\theta_{n}) = \mathbb{E}_{\mathcal{O},\boldsymbol{a}_n \sim \mathcal{M}_n} \;\;\;\;\;\;\;\;\;\;\;\;\;\; \;\;\;\;\;\;\;\;\;\;\;\;\;\;\;\;\;\;\;\;\;\;\;\;\;\;\;\; \\ \big[\nabla_{\theta_{n}} \vartheta_{n}(\boldsymbol{a}_{n}|\boldsymbol{o}_{n}) \nabla_{\boldsymbol{a}_{n}} Q_{n}^{\vartheta}(\mathcal{O}, \boldsymbol{a}_{1}, \dots, \boldsymbol{a}_{N})  |{\boldsymbol{a}_{n}=\vartheta_{n}(\boldsymbol{o}_{n})}\big],
\end{split}
\end{equation}
where according to Definition \ref{def:game_stage_game}, $\mathcal{M}_n$ represents the experiences for all the agents $(\mathcal{O}, \mathcal{O}',\boldsymbol{a}_{1}, \dots, \boldsymbol{a}_{N}, r_{1}, \dots, r_{N})$ which includes both the previous and current observations, $\mathcal{O}$ and $\mathcal{O}'$, respectively. The centralized action value function for all agents can be represented as follows \cite{IEEEhowto:Foerster_Communicate_Deep_multi_agent}:
\begin{equation} \label{eq:new_centralized_action_value}
\begin{split}
L(\theta_{n})  = \underset{\vartheta} \min \; \mathbb{E}_{\boldsymbol{o}_{n}, \boldsymbol{a}_{n}}\big[\frac{1}{2}(Q_{n}^{\vartheta}(\mathcal{O}, \boldsymbol{a}_{1}, \dots, \boldsymbol{a}_{N}) - y)^2\big],
\end{split}
\end{equation}
and
\begin{equation} \label{eq:new_y_value}
\begin{split}
y = r_{n} + \gamma Q_{n}^{\vartheta'}(\mathcal{O}', \boldsymbol{a}'_{1}, \dots, \boldsymbol{a}'_{N}) |{ \boldsymbol{a}'_{j}=\vartheta'_{j}(o'_{j})},
\end{split}
\end{equation}
where for parameters $\theta'_{n}$, the set of target polices is determined by $\mathcal{\vartheta'} = \left\{{\vartheta_{\theta'_{1}},\vartheta_{\theta'_{2}},\dots,\vartheta_{\theta'_{N}}}\right\}$. Therefore, to execute centralized action value function \eqref{eq:new_centralized_action_value}, the actions for all agents need to know, whereas the nature of environment is stationary as the policies are changing. In this scenario, we can efficiently learn the other agents' polices from the observations. Let us consider parameters $\phi$ for each agent $n$ that can maintain an approximation policy $\hat{\vartheta}_{\phi_{n}^{j}}$ from the observed policy $\vartheta_j$ of agent $j$. Thus, the loss function is defined as follows:    
\begin{equation} \label{eq:entroyp_centralized_action_value}
\begin{split}
L(\phi_{n}^{j})  = -\mathbb{E}_{\boldsymbol{o}_{j}, \boldsymbol{a}_{j}}\big[ \hat{\vartheta}_{n}^{j}(\boldsymbol{a}_{j}|\boldsymbol{o}_{j}) + \beta h(\hat{\vartheta}_{n}^{j})\big],
\end{split}
\end{equation}
where $\beta$ is a coefficient for the magnitude of regularization for solving the bias problem and $h(\cdot)$ determines the entropy for the policy distribution $\hat{\vartheta}$. Now, we can rewrite \eqref{eq:new_y_value} using this approximation, where $\hat{\vartheta_n}'$ determines the target policy networks for the policy $\hat{\vartheta_n}$ and redefined as follows \cite{IEEEhowto:Lowe_Multi_agent_actor_critic}:
\begin{equation} \label{eq:entroyp_new_y_value}
\begin{split}
y \approx \hat{y} = r_{n} + \gamma Q_{n}^{\vartheta'}(\mathcal{O}',\hat{\vartheta}'_{n}(\boldsymbol{o}_{1}), \dots, {\vartheta}'_{n}(\boldsymbol{o}_{n}), \dots \hat{\vartheta}'_{n}(\boldsymbol{o}_{N})).
\end{split}
\end{equation}
Hence, the objective function for the policy $\vartheta$ looks as follows:
\begin{equation} \label{eq:obj_deep_learning}
\begin{split}
L(\phi_{n})  = \underset{\vartheta_{\phi_{n}}} \min \; \mathbb{E}_{\boldsymbol{o}_{j}, \boldsymbol{a}_{j}}\big[\frac{1}{2}(Q_{n}^{\vartheta}(\mathcal{O}, \boldsymbol{a}_{1}, \dots, \boldsymbol{a}_{N}) - \hat{y})^2\big],
\end{split}
\end{equation}
and the policy gradient is as follows:
\begin{equation} \label{eq:final_gradient_policy_grdient}
\begin{split}
\nabla_{\phi_n} J(\phi_n) \approx \frac{1}{N} \mathbb{E}_{\mathcal{O},\boldsymbol{a}_n \sim \mathcal{M}_n} \;\;\;\;\;\;\;\;\;\;\;\;\;\;\;\;\;\;\;\;\;\;\;\; \;\;\;\;\;\;\;\;\;\;\;\;\;\;\;\;\;\;\;\;\;\;\;\;\\ \big[ \sum_{n \in N}
\nabla_{\phi_n} \log \vartheta_n(\boldsymbol{a}_{n}|\boldsymbol{o}_{n}) \nabla_{a_n} Q_{n}^{\vartheta}(\mathcal{O}, \boldsymbol{a}_{1}, \dots, \boldsymbol{a}_{N}) |_{{\boldsymbol{a}_{n}=\vartheta_{n}(\boldsymbol{o}_{n})}}\big].
\end{split}
\end{equation}

The Adaptive Moment Estimation (ADAM) optimizer has been widely used for function approximation \cite{IEEEhowto:adam_Kingma}. This method employs both first and second moments of the gradients and computes the individual adaptive learning rates for different batches of observations with different parameters. Hence, the first and second moment of gradients from \eqref{eq:obj_deep_learning} are as follows:
\begin{equation} \label{eq:first_moment}
\varrho_{t'}^{w} = \upsilon_1\varrho_{t}^{w} + (1-\upsilon_1) \nabla^w L(\phi_{n}), 
\end{equation}
\begin{equation} \label{eq:second_moment}
\nu_{t'}^{w} = \upsilon_2\nu_{t}^{w} + (1-\upsilon_2) (\nabla^w L(\phi_{n}))^2,
\end{equation}
where $\upsilon_1$ and $\upsilon_2$ are the decay rates. As a result, the ADAM optimizer can compute a bias correction (hyper-parameters correction) for both first and second order moments before a weight change calculation. This is important for the first few steps of training to tackle biasness. Thus, the corrected bias estimation functions are as follows:     
\begin{equation} \label{eq:bias_corrected_first_moment}
\hat{\varrho^{w}} = \frac{\varrho_{t'}^{w}}{1- (\upsilon_1)_{t'}}, 
\end{equation}
\begin{equation} \label{eq:bias_corrected_second_moment}
\hat{\nu^{w}} = \frac{\nu_{t'}^{w}}{1- (\upsilon_2)_{t'}},
\end{equation}
where the bias correction for first and second moments are determined by \eqref{eq:bias_corrected_first_moment} and \eqref{eq:bias_corrected_second_moment}, respectively. Therefore, the weight change $\Delta w_{t}$ of $L(\phi_{n})$ is defined  follows:
\begin{equation} \label{eq:change_weight}
\Delta w_{t} = - \ell \frac{\hat{\varrho^{w}}}{\sqrt{\hat{\nu^{w}}} + \kappa},
\end{equation}
where $\ell$ is the learning rate and a very small value of $\kappa$ prevents division by zero. As a result, the updated weight for the next time slot $t'$ is as follows:
\begin{equation} \label{eq:update_weight}
w_{t'} = w_t + \Delta w_{t}.
\end{equation}

To design shared neural networks for the proposed multi-agent A3C model for risk-aware energy scheduling, we use a rectified linear unit (ReLU) activation function \cite{IEEEhowto:ReLU_Nair}, which is able to handle nonlinearity and provides good approximations for various combinations of nonlinearity at a smaller computational cost. The ReLU activation function is defined as follows:
\begin{equation} \label{eq:rl_relu_act}
f(\boldsymbol{a}_t) = \max (0,\boldsymbol{a}_t),
\end{equation}
where, $\boldsymbol{a}_t$ is the action for energy storage and buying.    
\begin{algorithm}[t!]
	\caption{Risk-Aware Energy Scheduling MADRL Model Based on Asynchronous Advantage Actor-Critic (A3C)}
	\label{alg:Energy_Scheduling_Based_on_A3C}
	\begin{algorithmic}[1]
		\renewcommand{\algorithmicrequire}{\textbf{Input:}}
		\renewcommand{\algorithmicensure}{\textbf{Output:}}
		\REQUIRE  $\boldsymbol{s}_t = {( \mathcal{E}^{\textrm{dem}}(t), \mathcal{E}^{\textrm{ren}}(t), \mathcal{E}^{\textrm{sto}}(t), P(H_{\alpha}(\boldsymbol{a}_t, \xi)) )} \in \mathcal{S}$
		\ENSURE Trained Model: \textit{madrl}
		\\ \textbf{Initialization:}all agents $N$, $\gamma$, $\ell$, \textit{maxEpisodes}, $\mathcal{T}$, DQN 
		\FOR {\textbf{Until:} \textit{maxEpisodes}}
		\STATE \textbf{Constraints:} \eqref{Opt_1_1:const2},\eqref{Opt_1_1:const3} and \eqref{Opt_1_1:const4}
		\FOR {$\forall t \in \mathcal{T}$} 
		\FOR {$\forall n \in \mathcal{N}$}
		\STATE \textbf{Constraints:} \eqref{Opt_1_1:const5} and \eqref{Opt_1_1:const7}
		\STATE \textbf{Initialization:} $\boldsymbol{o}_{n} \coloneq ( \boldsymbol{s}_{t}, \boldsymbol{a}_{t}, r_{t}, \boldsymbol{s}_{t'} ) \in \mathcal{O}_n$
		\FOR{\textbf{Until:}\\$\mathbb{E}_{\boldsymbol{o}_n \in \mathcal{O}_n} \big[(Q^{\pi_{\theta_{n}}^*}(\boldsymbol{s}_{t},\boldsymbol{a}_{t}|\theta_{t}) - y_{t})^2\big]$ in eq. \ref{eq:loss_fn_dqn}}
		\STATE \textbf{Calculate:} $\underset{\boldsymbol{a}_{t} \in \mathcal{A}} \max \; \sum_{t'=t}^{\infty} \gamma^{t'-t} \mathbb{E}_{\pi_{\theta_{n}}} \big[r_{n}(\boldsymbol{a}_{t}, \boldsymbol{s}_{t})|\mathcal{O}_n\big]$ in eq. \eqref{eq:game_dis_count_state_value}
		\STATE \textbf{Using:} eq. \eqref{eq:update_weight}, \eqref{eq:rl_relu_act}, and \eqref{eq:rl_soft_max}
		\STATE \textbf{Action:}  $\boldsymbol{a}_t \sim \pi_{\theta}(\boldsymbol{a}_t|\boldsymbol{s}_t)$
		\STATE \textbf{Receive:} $\boldsymbol{o}_n=(\boldsymbol{s}_{t}, \boldsymbol{a}_{t}, r_{t}, \boldsymbol{s}_{t'})$
		\STATE \textbf{Evaluate:} $L(\phi_{n})$ using eq. \eqref{eq:obj_deep_learning}
		\STATE \textbf{Calculate:} $\nabla_{\phi_n} J(\phi_n)$ using eq. \eqref{eq:final_gradient_policy_grdient} 
		\STATE \textbf{Update:} $\theta_{n} = \theta_{n} + \nabla_{\theta_{n}} J(\theta_{n})$
		\ENDFOR
		\STATE \textbf{Update:} $\phi_n = \phi_n + \nabla_{\phi_n} J(\phi_n)$
		\STATE \textbf{Calculate:} $\nabla_{\theta_{n}} J(\theta_{n})$ using eq. \eqref{eq:new_gradient_policy_grdient} 
		\ENDFOR
		\STATE \textbf{Append:} $\boldsymbol{o}_t \in \mathcal{O}$
		\STATE \textbf{Update:} Policy $\pi_{\theta_{n}}$, Value $V^{\pi_{\theta_{n}}}(\boldsymbol{s}_{t})$
		\ENDFOR
		\STATE \textbf{Update MADRL model:} \textit{madrl} 
		\ENDFOR
		\RETURN \textit{madrl}
	\end{algorithmic} 
\end{algorithm}
To determine the output from the neural networks, in this model we use the Softmax activation function \cite{IEEEhowto:softmax_Sutton}, which is appropriate for a cross-entropy cost function \eqref{eq:entroyp_centralized_action_value}. This function has the properties of a negative log probability and a very large gradient, which is suitable for estimating the gradient \eqref{eq:final_gradient_policy_grdient} of problem \eqref{Opt_2_1}. The Softmax function is defined as follows:
\begin{equation} \label{eq:rl_soft_max}
P(\boldsymbol{a}_t) = \frac{L(\phi_{n})/\tau)}{\sum_{\forall p \in P}L(\phi_{n})/\tau)}, 
\end{equation}
where $L(\phi_{n})$ is the cost function from \eqref{eq:obj_deep_learning}, $\tau$ determines the temperature parameter, and $P$ is the number of activated neurons. For a large value of $\tau \to \infty$, $P(\boldsymbol{a}_t)$ is near to zero and a lower value of $\tau$ provides the highest expectation of the action probability $P(\boldsymbol{a}_t)$ (tends to 1).

The proposed risk-aware energy scheduling MADRL model in Algorithm 
\ref{alg:Energy_Scheduling_Based_on_A3C}, which is run by the microgrid controller. Additionally, the microgrid controller receives the necessary information regarding energy demand of each time slot intervals of the MEC network from the MBS. Therefore, this algorithm verifies constraints from \eqref{Opt_1_1:const2} to \eqref{Opt_1_1:const7} (represents as \eqref{Opt_2_1:const2} in problem \eqref{Opt_2_1}) from lines 2 to 5, where line 2 ensures constraints \eqref{Opt_1_1:const2}, \eqref{Opt_1_1:const3} and \eqref{Opt_1_1:const4}. Line 5 provides an assurance of binary decision (storing/buying) ${a}_t \in \left\lbrace0,1 \right\rbrace$ for each time slot $t$ in the time horizon $\mathcal{T}$. The DQN of the proposed MADRL is developed through lines 7 to 15, where each agent $n$ (the actor) calculates the loss \eqref{eq:loss_fn_dqn} (in line 7) using the DQN and evaluates by critic \eqref{eq:obj_deep_learning} (in line 12) to update the gradient of the loss function \eqref{eq:final_gradient_policy_grdient} (in line 13). Therefore, the weight of local policy updates in line 16 and the gradient of the global agent \eqref{eq:new_gradient_policy_grdient} is determined in line 17. Finally, the observation of each time slot $\boldsymbol{o}_t$ is appended into the observational set $\mathcal{O}$ (in line 19) and updates the parameterized policy $\pi_{\theta_{n}}$ and value $V^{\pi_{\theta_{n}}}(\boldsymbol{s}_{t})$ (in line 20) for the further use of the MADRL model. Hence, Algorithm \ref{alg:Energy_Scheduling_Based_on_A3C} provides a risk-aware energy scheduling MADRL model (in line 24) for the microgrid-powered MEC network.
\begin{algorithm}[t!]
	\caption{Risk-Aware Energy Scheduling Forecasting based on the MADRL Model}
	\label{alg:Scheduling}
	\begin{algorithmic}[1]
		\renewcommand{\algorithmicrequire}{\textbf{Input:}}
		\renewcommand{\algorithmicensure}{\textbf{Output:}}
		\REQUIRE   MADRL model: \textit{madrl}, current state: $\boldsymbol{s}_t$ 
		\ENSURE  $\boldsymbol{a}_{t'}$, $\boldsymbol{s}_{t'}$ 
		\\ \textbf{Initialization}: \textit{model} = \textit{loadweights}(\textit{madrl}), \textit{done} =$100$
		\FOR {\textbf{Until:} \textit{done}}
		\STATE \textit{policy = model($\boldsymbol{s}_t$)}
		\STATE \textbf{Calculate:} \textit{policy} using eq. \eqref{eq:rl_soft_max}
		\STATE $\boldsymbol{a}_{t'}$ = $\argmax (\textit{policy})$
		\STATE $\boldsymbol{s}_{t'}$ = CalForecast($\boldsymbol{a}_{t'}$)
		\STATE \textbf{Update:} Policy ${\pi_{\theta}}_n$, Value $V^{\pi_{\theta_{n}}}(\boldsymbol{s}_{t'})$  
		\ENDFOR
		\STATE \textbf{Update MADRL model:} \textit{model}
		\RETURN $\boldsymbol{a}_{t'}$, $\boldsymbol{s}_{t'}$  
	\end{algorithmic} 
\end{algorithm}

The convergence of Algorithm \ref{alg:Energy_Scheduling_Based_on_A3C} is discretized via $N$-agent RL settings, where we consider an action space $\boldsymbol{a}_{tn} \in \mathcal{A}$ with two actions $\zeta_t^1$ and $\zeta_t^0$ at time slot $t$. To determine the gradient step, we use a probabilistic model, where the gradient step moves toward the correct direction and decreases exponentially with an increasing number of agents. We investigate the convergence via the following Proposition: 
\begin{proposition}
	\label{pro:Convergence_Multi_Agent_Risk_Sensitive}
		Consider an unknown environment with state space $\boldsymbol{s}_{t} \in \mathcal{S}$ with $N$ agents such that all agents are initialized an equal probability of $\frac{1}{2}$ for the binary actions, $P(\boldsymbol{a}_{n} = \zeta_t^1) = \theta_{n} = \frac{1}{2}, \forall n \in \mathcal{N}$, where $r_{n}(\boldsymbol{a}_{1}, \dots, \boldsymbol{a}_{N}) = 1 |(\boldsymbol{a}_{1}= \dots = \boldsymbol{a}_{N}) $ . If we estimate the gradient $\hat{\nabla}_{\theta_{n}} J(\theta_{n})$ of the cost function \eqref{eq:obj_policy_grdient}, then we get the following relationship with the true gradient $\nabla_{\theta_{n}} J(\theta_{n})$: 
		\begin{equation} 
		\label{eq:Convergence_relation_policy_grdient}
		\begin{split}
		P\left((\hat{\nabla}_{\theta_{n}} J(\theta_{n}), \nabla_{\theta_{n}} J(\theta_{n})) > 0\right) \propto \left(\frac{1}{2}\right)^N.
		\end{split}
		\end{equation}
[See Appendix \ref{apd:Convergence_of_Proposed_Model}]. 
\end{proposition}
Proposition \ref{pro:Convergence_Multi_Agent_Risk_Sensitive} justifies that for a single observation the proposed multi-agent risk-aware energy scheduling model achieves convergence, which implies that this model is able to converge toward the multiple observations.

To forecast the risk-aware energy scheduling of the microgrid-powered MEC network, we propose Algorithm \ref{alg:Scheduling}, where this algorithm uses the trained model of Algorithm \ref{alg:Energy_Scheduling_Based_on_A3C}. In Algorithm \ref{alg:Scheduling}, line 2 decomposes the trained MADRL model for the current state $\boldsymbol{s}_t$. An action (buying/storing) $\boldsymbol{a}_{t'}$ is taken through the lines 3 and 4 for the next state $\boldsymbol{s}_{t'}$ (energy scheduling) forecasting. Line 5 determines the next state information $\boldsymbol{s}_{t'}$ based on the action $\boldsymbol{a}_{t'}$, which includes the MEC energy demand $\mathcal{E}^{\textrm{dem}}(t')$, renewable energy generation $\mathcal{E}^{\textrm{ren}}(t')$, storage energy $\mathcal{E}^{\textrm{sto}}(t')$, and the CVaR risk for next time slot $t'$. 

We analyze CVaR and make a reflection on the learning model. To do this, we generate the state-space in such a way that the distribution of CVaR is considered. Therefore, we calculate CVaR by applying a normal distribution and embed it into state information before anticipating the learning model, where the computational complexity belongs to $O(S^3)$. In order to accelerate the learning process with a low complexity deep learning model, we design a policy network in shared neural networks (weight sharing) manner among the virtual agents and global agent. The goal of all virtual agent $N$ is always the same and the policy gradient increases linearly with respect to the number of iteration (i.e., the total number of weight $\vartheta_n$ updates in each time slot $t$). Thus, the overall computational complexity of the policy networks leads to $O(|\mathcal{S}|^2|\mathcal{A}||\mathcal{N}|)$ \cite{IEEEhowto:Kaelbling_Complexity_Policy}, where a single virtual agent complexity goes in $O(|\mathcal{S}|^2|\mathcal{A}|)$ since learning time is decreasing \cite{ IEEEhowto:Whitehead_MRL_Complexity_Base} at a rate $O(\frac{1}{n}), \forall n\in \mathcal{N}$. The experimental analysis and insightful discussion of the risk-aware energy scheduling are given in the later section.  

\section{Experimental Analysis and Discussion}
\begin{table}[t!]
	\caption{Summary of Experiment Setup }
	\begin{center}
		\begin{tabular}{|c|c|}
			\hline
			\textbf{Description}&{\textbf{Value}} \\
			\hline
			No. of SBSs &10\\ 
			\hline
			No. of servers in each SBS &5\\
			\hline	
			No. of CPU cores in one server&4 with 1.2 GHz \cite{IEEEhowto:Raspberry_energy, IEEEhowto:Meng_pi_use_edge}\\
			\hline
			No. of solar units &40 \cite{IEEEhowto:UMass_Solar_panel_dataset}\\
			\hline	
			Task sizes &[31,1546060] bytes \cite{IEEEhowto:CRAWDAD_dataset_nyupoly_video}\\
			\hline
			One time slot $t$ &15 minutes \cite{IEEEhowto:Munir_Edge_Microgrid}\\
			\hline
			No. of tasks request at each SBS & [1,10000] \cite{IEEEhowto:Munir_Edge_Microgrid}\\
			\hline
			CVaR confidence levels &$[90\%, 95\%, 99\%]$\\ 
			\hline				
			Maximum number of episodes & $1000$\\
			\hline
			No. of agents $N$ & $[4, 8, 10]$\\
			\hline
			Learning rate $\ell$ & $10^{-3}$\\
			\hline	
			Discount factor $\gamma$ &$0.99$\\
			\hline
			No. of hidden layers and neurons &$2$, $100$\\
			\hline	
		\end{tabular}
		\label{tab2_sim_param}
	\end{center}
	\vspace{-6mm}
\end{table}
In this section, we evaluate the proposed model using extensive experimental analyses. We implement the \emph{risk-aware energy scheduling} model via Python, along with TensorFlow APIs. We run the simulated model using a core i$7$ processor with a speed of $2.6$ GHz along with $8$ GB of RAM as a microgrid controller. 

To evaluate this model, we used the well-known UMass solar panel dataset \cite{IEEEhowto:UMass_Solar_panel_dataset} for renewable energy generation information, as well as the CRAWDAD nyupoly/video dataset \cite{IEEEhowto:CRAWDAD_dataset_nyupoly_video}, for estimating the energy consumption of the MEC networks. In particular, we can use UMass solar panel dataset \cite{IEEEhowto:UMass_Solar_panel_dataset} as microgrid energy generation since solar panels are installed in the same geographical location \cite{IEEEhowto:Solar_Panels_UMASS}. We consider the network parameters are same as the dataset in \cite{IEEEhowto:CRAWDAD_dataset_nyupoly_video}.  Further, we divided both datasets into $70\%$ and $30\%$ for training and testing, respectively \cite{IEEEhowto:Munir_Edge_Microgrid}. 
In addition, we prepossess both datasets to extract the state-space information $\forall \boldsymbol{s}_t \coloneq {( \mathcal{E}^{\textrm{dem}}(t), \mathcal{E}^{\textrm{ren}}(t), \mathcal{E}^{\textrm{sto}}(t), P(H_{\alpha}(\boldsymbol{a}_t, \xi)) )} \in \mathcal{S}$, where we determine the value of $P(H_{\alpha}(\boldsymbol{a}_t, \xi))$ using Algorithm \ref{alg:CvaR}, and Table \ref{tab2_sim_param} describes the important parameters of the experiment setup. To the best of our knowledge, the literature does not provide similar analogy of the risk-aware energy scheduling for microgrid-powered MEC networks. However, to provide more concrete results 
\footnote{In this work, we  have shown the experimental results based on the accuracy performance metric of the proposed model from the dataset. If we considered high capacity MEC hosts (i.e., range of $3.3$ GHz to $20$ GHz)
\cite{IEEEhowto:Sun_EMM,IEEEhowto:Li_MEC_Capacity_3_3GHz, IEEEhowto:Li_MEC_Capacity_4GHz_to_8GHz, IEEEhowto:Tran_MEC_Capacity_20GHz, IEEEhowto:Guo_MEC_Capacity_800M_to_4GHz} in the experimental environment, the proposed model still can perform energy consumption estimation for that MEC hosts without any side-effect. Moreover, the numerical value of the energy consumption will be changed based on Thermal Design Power (TDP) and other parameters of the MEC host. The training and testing performance of the proposed multi-agent model will remain the same due to the execution of model without any internal/algorithmic changes. In fact, we only need to provide the new dataset as an input for those MEC hosts.}, we compared our proposed MADRL model with single agent A2C and random-agent A3C model as baseline models. 
In particular, we choose neural advantage actor-critic (A2C) \cite{IEEEhowto:single_agent_Sutton_RL_book, IEEEhowto:AC_Base_Takahashi} and asynchronous advantage actor-critic (A3C) based random-agent deep reinforcement learning (DRL) framework \cite{IEEEhowto:Lowe_Multi_agent_actor_critic} as two baselines to provide a fair and effective comparison analysis with the proposed multi-agent A3C model. In which, neural A2C plays the role of a single-agent (i.e., centralized) solution and the random-agent A3C model is considered as a distributed scheme, where we have deployed the same neural architecture and setting with the proposed A3C-based MADRL model. In fact, we have proposed the model in such a way that the decision of energy store/buying comes from the current state. Therefore, neural A2C (i.e., single agent), and random-agent A3C are capable of playing the role of strong baselines for effective and fair comparison of the proposed MADRL model. Thus, the learning environment of neural A2C \cite{IEEEhowto:single_agent_Sutton_RL_book, IEEEhowto:AC_Base_Takahashi} encompasses the entire energy environment (i.e., state information) of the considered microgrid-powered MEC network. Meanwhile, the random-agent A3C can find its own policy by centralized training with decentralized execution, where the environment (i.e., state information) remains the same for all of the random agents.
Additionally, we have compared a tail-risk of the energy shortfall by considering both normal and Student's t distribution for VaR and CVaR. In the experiment, the \emph{actual value} of energy generation decision represents the true value of the decision variable in \eqref{eq:zeta_decision_variable} $a_t$ (i.e., action) for each time slot $t$. Where if $a_t = 0$ then we need to buy energy from the main grid, and $a_t = 1$ that indicates the store decision. This \emph{actual value} decision comes from the datasets. Therefore, \emph{ground truth} considered as an optimal decision for the proposed scenario. In fact, we consider the network parameters are the same as the dataset in \cite{IEEEhowto:CRAWDAD_dataset_nyupoly_video}. From \cite{IEEEhowto:UMass_Solar_panel_dataset} and \cite{IEEEhowto:CRAWDAD_dataset_nyupoly_video} datasets we get the \emph{actual value} of energy generation and consumption, respectively.
\begin{figure}
	\begin{subfigure}{.49\textwidth}
		\centering
		\includegraphics[width=\textwidth]{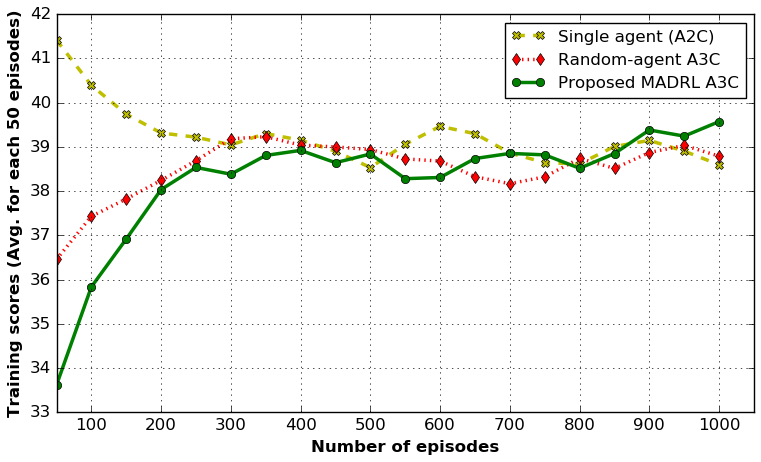}
		\caption{Training reward}
		\label{fig:Training_reward}
	\end{subfigure}%
	\\
	\begin{subfigure}{.49\textwidth}
		\centering
		\includegraphics[width=\textwidth]{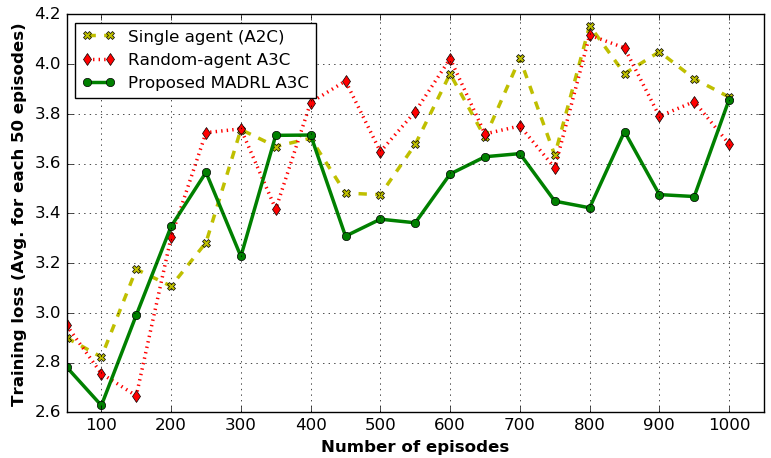}
		\caption{Training loss}
		\label{fig:Training_Loss}
	\end{subfigure}
	\caption{Illustration of training scores (i.e., rewards) and losses for 95\% CVaR confidence.}
	\label{fig:Training_score}
	\vspace{-6mm}
\end{figure}
\begin{figure}[!t]
	\centering
	\includegraphics[width=\linewidth]{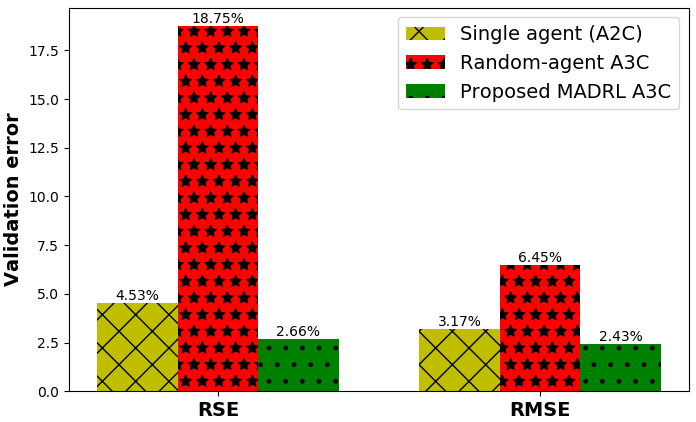}
	\caption{Training validation error analysis for the proposed risk-aware energy scheduling for 95\% CVaR confidence.}
	\label{fig:validation_error}
	\vspace{-6mm}
\end{figure} 
\begin{figure*}[!h]
	\centering
	\includegraphics[width=15cm]{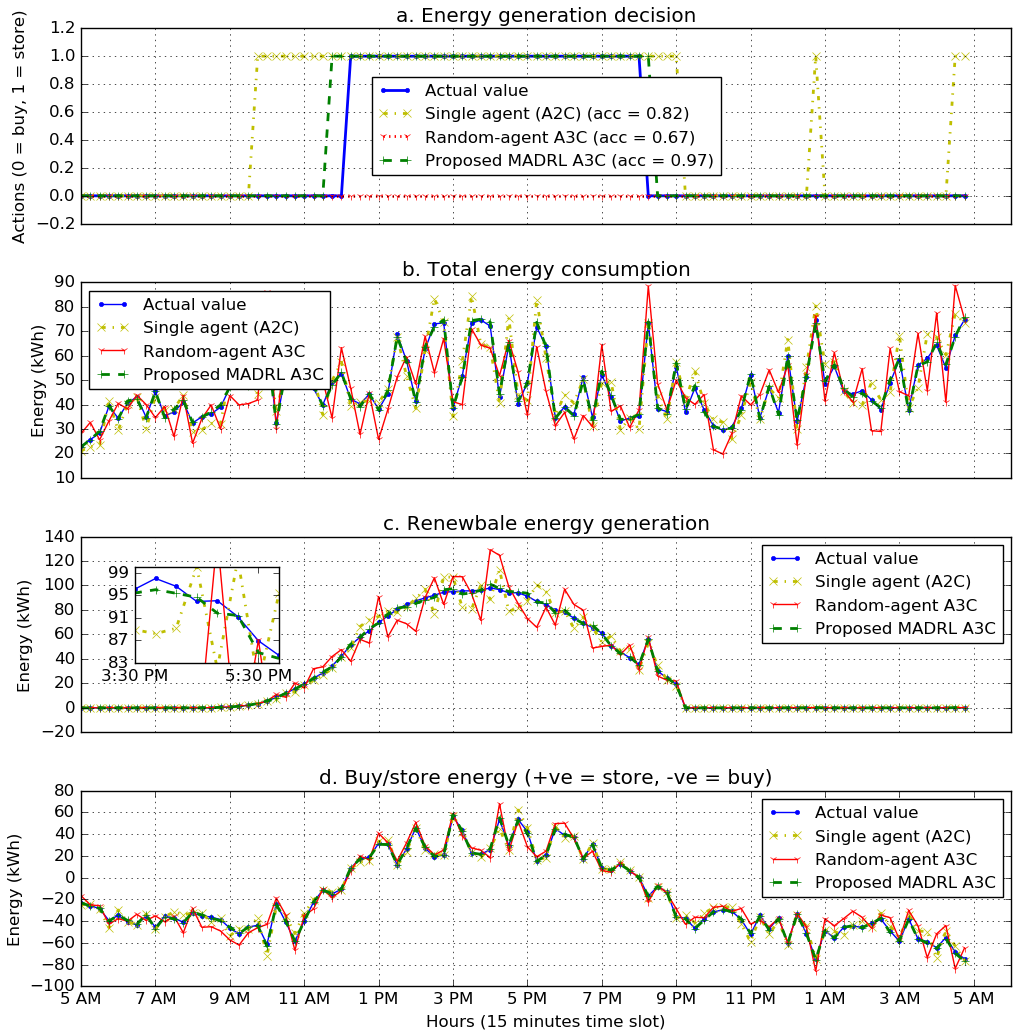}
	\caption{Risk-aware energy scheduling training model validation with $95\%$ CVaR confidence for one day with a $15$-minute time-slot.}
	\label{fig:profile_training}
	\vspace{-6mm}
\end{figure*}
 \begin{figure}[!t]
	\centering
	\includegraphics[width=8.9cm]{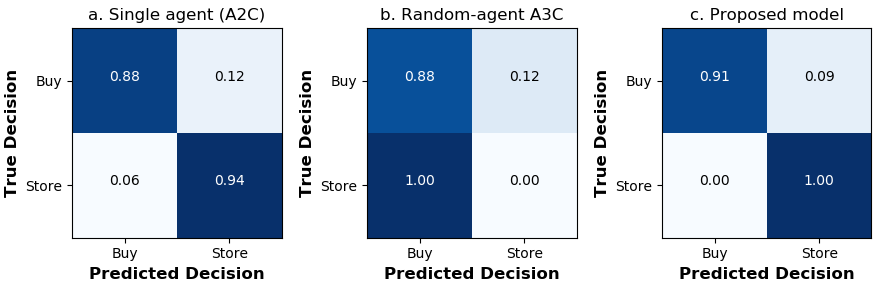}
	\caption{Confusion matrix testing with $95\%$ CVaR confidence for model test validation.}
	\label{fig:confusion_matrix_testing}
	\vspace{-4mm}
\end{figure}
\begin{figure}[!t]
	\centering
	\includegraphics[width=8.9cm]{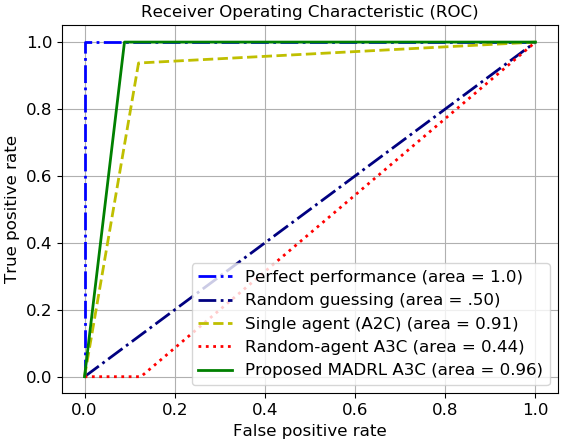}
	\caption{Receiver operating characteristic (ROC) curve for test validation with $95\%$ CVaR confidence.}
	\label{fig:ROC_testing}
	\vspace{-6mm}
\end{figure}
\begin{figure}[!t]
	\centering
	\includegraphics[width=8.9cm]{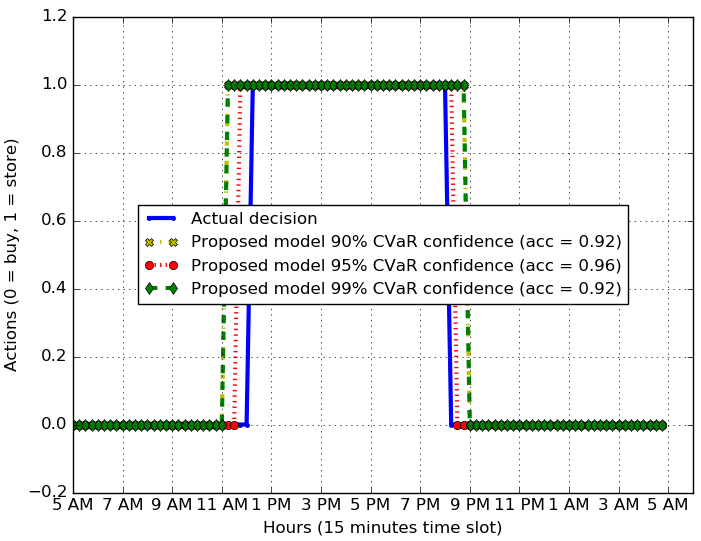}
	\caption{Risk-aware energy scheduling model testing using CVaR confidence levels of $90\%$, $95\%$, and $99\%$.}
	\label{fig:Risk_Profile_Testing}
	\vspace{-4mm}
\end{figure}
\subsection{MADRL-based Risk-Aware Energy Scheduling Training Validation}
In order to validate the training model, first we analyze the convergence of the proposed MADRL Algorithm \ref{alg:Energy_Scheduling_Based_on_A3C} and compare it with the single agent A2C and random-agent A3C models. Training scores and losses for the single agent A2C (cross mark with a dashed line), random-agent A3C (diamond mark with a dotted line), and  proposed multi-agent asynchronous A3C (circle mark with a solid line) models are shown in Fig. \ref{fig:Training_score}. Reward in Fig. \ref{fig:Training_reward} illustrates the convergence of proposed model with higher score than the other two models. In Fig. \ref{fig:Training_reward}, at the beginning of training, single-agent (deep actor-critic) has achieved higher score than the multi-agent due to the less variation among the training dataset. In particular, when the randomness appears for the input dataset, the training score of the single-agent method gradually decreases. In fact, due to the less variation among the exploration and exploitation of the single-agent model, it cannot infer the uncertainty when an energy environment (i.e., generation and consumption) is unknown. On the other hand, in case of multi-agent, at the beginning of training it achieves lower reward than a single agent model due to the more variation among exploration and exploitation by each agent. In general\cite{IEEEhowto:He_Green_RL, IEEEhowto:softmax_Sutton}, single-agent mechanism is not capable of choosing an action for the best policy due to limited information and the single-agent reinforcement learning only optimizes the action policy for itself only. However, by changing the environment this method cannot cope with an unknown environment due to the diversity. In our proposed model, each agent learns action policy based on diverse feedbacks from the other agents as well as its own state-space. In particular, the global agent (i.e., critic) of the proposed MADRL A3C model can optimizes a decision toward the best energy scheduling policy for the microgrid-powered MEC network. 
Fig. \ref{fig:Training_Loss} shows that from episodes $601$ to $1000$ the average losses of the single-agent deep RL (A2C), random agent deep RL (A3C) and proposed multi-agent A3C-based deep reinforcement learning methods are $3.92$, $3.84$, and, $3.58$, respectively. In particular, the proposed MAMRL method achieves $9.52\%$ and $7.28\%$ less training loss as compared to single-agent and random-agent models, respectively, between training episodes $601$ and $1000$. Further, Fig. \ref{fig:Training_Loss} shows that the more fluctuation of the training losses for the single-agent and random-agent models occur between episodes $751$ and $800$, in which the percentage of training losses for a single agent and random agent models are $12.53\%$ and $12.96\%$, respectively. However, the training loss of the proposed model is $0.78\%$ between episodes $751$ and $800$. As a result, the training losses (in Fig. \ref{fig:Training_Loss}) of the proposed model is more stable than the others when an energy environment is changing due to uncertain energy generation and demand by the microgrid-powered MEC network.

Second, we present the training validation errors of the proposed MADRL model along with the other two baseline models in Fig. \ref{fig:validation_error}. We consider two performance metrics, Root Mean Square Error (RMSE), and Relative Squared Error (RSE) \cite{IEEEhowto:Radosavljevic_RSE} that can efficiently justify the validation error for a discrete outcome of the training model like our action (buy/storage). Fig. \ref{fig:validation_error} illustrates that the proposed MADRL scheme can achieve around $2\%$ more performance gain as compared with the single-agent method for the RSE validation metric. Here, the random-agent A3C model achieves a high error because the agents are executed randomly so that the policy does not cope with an uncertain environment, whereas the proposed model can deal with an uncertain environment by its individual agent.

Third, we illustrate one-day energy scheduling with a $15$-minute duration time slot for training model validation (in Fig. \ref{fig:profile_training}) with a $95\%$ CVaR confidence level, in which the action selection accuracy for the energy storing and buying decision of the proposed model gain around $97\%$, which is a feasible outcome in terms of training validation. However, the single agent and random agents models have accuracies of $82\%$ and $67\%$, respectively. Although the number of training episodes ($1,000$) is the same for all of the three models. Further, the random agent model cannot make the right decision with respect to energy storage due to more variation among the energy demand and generation over the time slots during the testing. Therefore, the random agent model cannot handle uncertain energy environment when demand and generation random over time. Thus, the proposed MADRL model achieves a higher accuracy than that the others due to the nature of information (observations) sharing among the agents of Algorithm \ref{alg:Energy_Scheduling_Based_on_A3C}.  
\begin{figure}
	\begin{subfigure}{.5\textwidth}
		\centering
		\includegraphics[width=\textwidth]{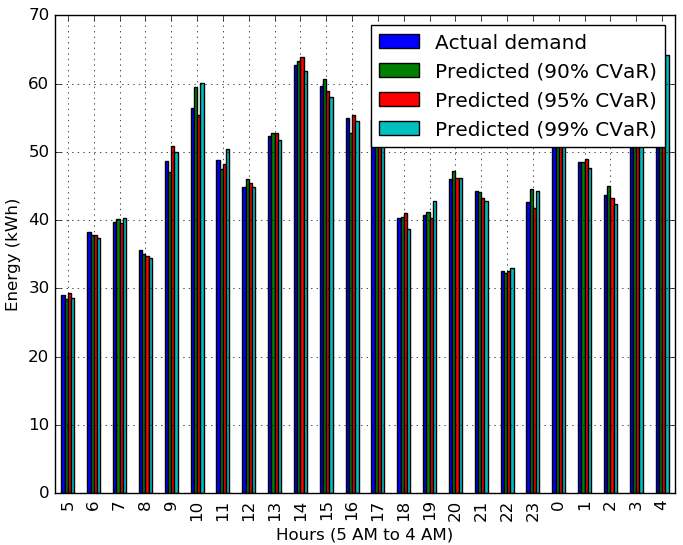}
		\caption{Total energy consumption}
		\label{fig:Risk_Profile_Testing_demand}
	\end{subfigure}%
	\\
	\begin{subfigure}{.5\textwidth}
		\centering
		\includegraphics[width=\textwidth]{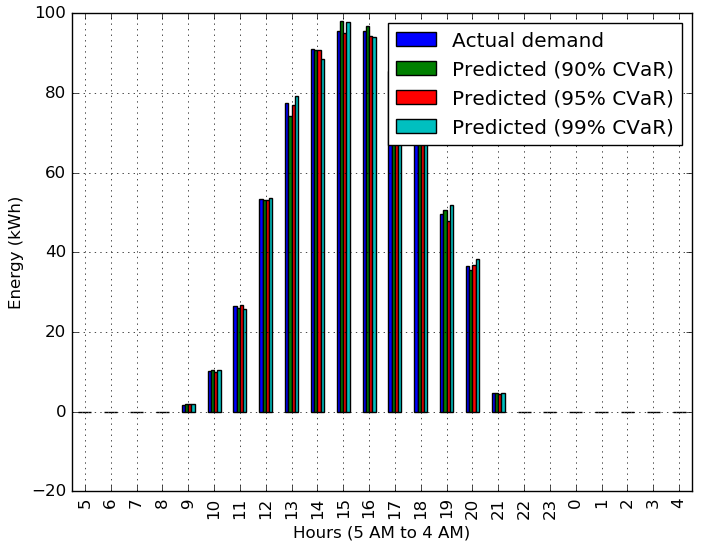}
		\caption{Renewbale energy generation}
		\label{fig:Risk_Profile_Testing_renewable}
	\end{subfigure}
	\\
	\begin{subfigure}{.5\textwidth}
		\centering
		\includegraphics[width=\textwidth]{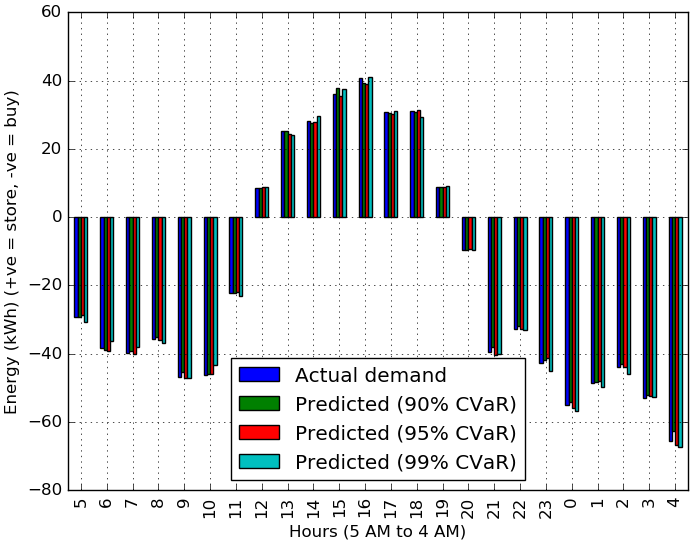}
		\caption{Buy/store energy (+ve = store, -ve = buy)}
		\label{fig:Risk_Profile_Testing_storage}
	\end{subfigure}
	\caption{Risk-aware energy forecast with CVaR confidence levels of $90\%$, $95\%$, and $99\%$ using the test dataset.}
	\label{fig:testing_profile_forecast}
\end{figure}
\begin{figure*}[!t]
	\centering
	\includegraphics[width=15cm]{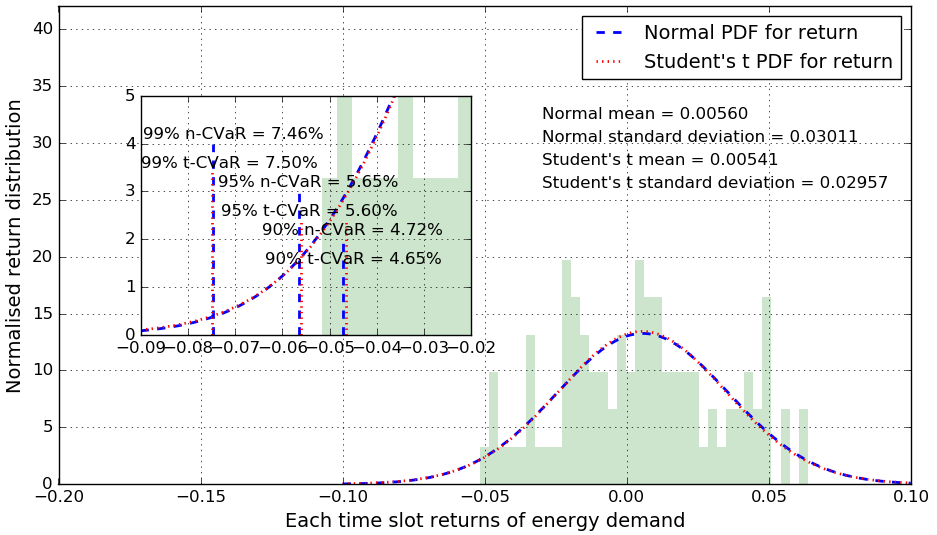}
	\caption{Conditional value-at-risk (CVaR) analysis for proposed risk-aware energy scheduling model.}
	\label{fig:CVaR_Risk}
	\vspace{-6mm}
\end{figure*}
\subsection{MADRL-based Risk-Aware Energy Schedule Testing Validation}
To illustrate the performance of the trained model, first we analyze the confusion matrix and receiver operating characteristic (ROC) as shown in Fig. \ref{fig:confusion_matrix_testing} and Fig. \ref{fig:ROC_testing}, respectively. Fig. \ref{fig:confusion_matrix_testing} represents the confusion matrix of test result validation with $95\%$ CVaR confidence, where the proposed model makes $9\%$ incorrect decision to determine the energy buying action (in Fig. \ref{fig:confusion_matrix_testing}). However, both the single-agent (in Fig. \ref{fig:confusion_matrix_testing}) and random-agents (in Fig. \ref{fig:confusion_matrix_testing}) models choose $12\%$ of incorrect actions for energy buying decision. On the other hand, in the case of storing decisions of the random-agents model, all ($100\%$) are incorrect decisions. Thus, confusion matrix (in Fig. \ref{fig:confusion_matrix_testing}) illustrates the significant perform gain of the proposed MADRL model under nondeterministic environment of microgrid-powered MEC networks. In Fig. \ref{fig:ROC_testing}, the random-agent A3C encompasses below $50\%$ for energy storing/buying decisions, while the single agent A3C covers around $91\%$ area of the ROC curve. However, the proposed model captures around $96\%$ of the area in the ROC curve, which assures that the proposed model performs significantly better than the other models.

Second, we validate the forecasting accuracy of Algorithm \ref{alg:Scheduling}, where Fig. \ref{fig:Risk_Profile_Testing} demonstrates the action selection of the one-day energy supply plan with a $15$-minute duration for CVaR confidence levels of $90\%$, $95\%$, and $99\%$. This figure illustrates that the testing accuracy is $92\%$, $96\%$, and $92\%$ for CVaR confidences of $90\%$, $95\%$, and $99\%$, respectively. Fig. \ref{fig:testing_profile_forecast} illustrates the overall energy profiling of the considered microgrid-powered MEC network using the proposed model for $24$ hours. In particular, Fig. \ref{fig:testing_profile_forecast} represents the total amount of energy consumption forecasting using CVaR confidences of $90\%$, $95\%$, and $99\%$. Further, Fig. \ref{fig:testing_profile_forecast} also presents the energy forecast for renewable and stored energy, respectively. The negative values of the stored energy are determined by the amount of energy that needs to be bought from the main grid at that time slot. To this end, the difference between forecasted and actual energy is negligible (in Fig. \ref{fig:testing_profile_forecast}), since for $95\%$ CVaR confidence the forecasting accuracy achieves around $96\%$.  

Finally, we investigate the tail of the CVaR in Fig. \ref{fig:CVaR_Risk} of the proposed risk-aware energy scheduling model. We analyze two types of distribution, the first is normal and the second is the student's t-distribution. The normal CVaR distribution is represented as blue dashed line, where we observe CVaR values of $4.72\%$, $5.65\%$, and $7.46\%$ with confidence levels of $90\%$, $95\%$, and $99\%$ (zoomed-in part in Fig. \ref{fig:CVaR_Risk}), respectively. Consequently, the student's t-distribution (red dotted line) achieves CVaR values of $4.65\%$, $5.60\%$, and $7.50\%$ with the confidences levels of $90\%$, $95\%$, and $99\%$ (zoomed-in part in Fig. \ref{fig:CVaR_Risk}), respectively. This analogy infers that the differences in CVaR values are very small between the two distributions, which validates the proposed risk-aware energy scheduling model under uncertainties. Additionally, this model significantly reduces the risk of energy shortfall of the microgrid-powered MEC network.

\section{Conclusions}
In this paper, we have introduced a conditional value-at-risk based energy scheduling model for a microgrid-powered MEC network using an $N$-agent stochastic game. We have mitigated the issue of volatilities for both the wireless network's energy consumption and the microgrid generation while considering uncertainties between demand and supply. Furthermore, we have achieved a joint policy Nash equilibrium, which determines the optimal energy scheduling policy of the proposed model. We solve this model by applying a multi-agent deep reinforcement learning approach, where we have admitted a shared neural networks with the asynchronous advantage actor-critic algorithm, thus ensuring high-accuracy energy scheduling with fast execution. Our extensive experimental results demonstrate a significant performance gain of the proposed approach, with this model providing upto $96\%$ accurate energy scheduling with $5.65\%$ risk for a $95\%$ CVaR confidence level, as compared with the single-agent and random-agents A3C models. Finally, our experimental results have established a risk-aware sustainable MEC network with respect to energy consumption and generation.

\appendices
\section{Proof of Proposition \ref{pro:discounted_reward_game_Nash_equilibrium_point} } \label{apd:discounted_reward_game_Nash_equilibrium_point_proof}
\begin{proof}
	For agent $n$, policy $\pi_{\theta_{n}}^*$ is the best response for the equilibrium responses from all other agents. The agent $n$ is unable to improve reward (payoff) $V^{\pi_{\theta_{n}}}(\boldsymbol{s}_{t}, \pi_{\theta_{n}}^*)$ by deviating from policy  $\pi_{\theta_{n}}^*$. Using \eqref{eq:game_Q_function}, we can write the following: 
	\begin{equation} \label{eq:discounted_reward_game_Nash_equilibrium}
	\begin{split}
	V^{\pi_{\theta_{n}}}(\boldsymbol{s}_{t}, \pi_{\theta_{n}}^*) \ge 	r_{n}(\boldsymbol{s}_{t}, \boldsymbol{a}_{t1} , \dots,\boldsymbol{a}_{tN}) \; + \;\;\;\;\;\;\;\;\;\;\;\;\;\;\; \;\;\;\;\;\;\;\;\;\; \\ \sum_{\boldsymbol{s}_{t'} \in \mathcal{S}, t'=t}^{\infty} \gamma^{t'-t} P(\boldsymbol{s}_{t'}|\boldsymbol{s}_{t}, \boldsymbol{a}_{t1}, \dots,\boldsymbol{a}_{tN}) \; V^{\pi_{\theta_{n}}}(\boldsymbol{s}_{t'}, \pi_{\theta_{1}}^*,\dots, \pi_{\theta_{N}}^*).
	\end{split}
	\end{equation} 
	Therefore, according to Definition \ref{def:game_stage_game}, the $N$-agent (player) stochastic game $\mathcal{G}$ is a multi-periods stage game, which has the properties of a joint strategy Nash equilibrium. Now, using \eqref{eq:game_joint_strategy_equilibrium_const}, we have
	\begin{equation} 
	\label{eq:_game_Nash_equilibrium_stage_stocastic}
	\begin{split}
	V^{\pi_{\theta_{n}}}(\boldsymbol{s}_{t}, \pi_{\theta_{n}}^*) = \vartheta_{\theta_{n}} \vartheta_{\theta_{\bar{n}}} {M}_n(\mathcal{O}_n, \boldsymbol{a}_{t}).
	\end{split}
	\end{equation} 
	Equation \eqref{eq:_game_Nash_equilibrium_stage_stocastic}  implies the inequality of \eqref{eq:game_joint_strategy_equilibrium_const}, which shows that $\pi_{\theta_{n}}^*$ is the equilibrium
	point of single-stage game and is denoted as follows:
	\begin{equation} 
	\label{eq:game_Nash_equilibrium_stage_prooved}
	\begin{split} 
	V^{\pi_{\theta_{n}}}(\boldsymbol{s}_{t}, \pi_{\theta_{n}}^*) \ge \hat{\vartheta}_{\theta_{n}} \vartheta_{\theta_{\bar{n}}} {M}_n(\mathcal{O}_n, \boldsymbol{a}_{t}), \forall \vartheta_{\theta_{n}}\in \hat{\vartheta}_{\theta_{n}} (\mathcal{A}_n).
	\end{split}
	\end{equation}
\end{proof}
\section{Proof of Proposition \ref{pro:Convergence_Multi_Agent_Risk_Sensitive}} \label{apd:Convergence_of_Proposed_Model}
	\begin{proof}
	Here, we can write the probability of action $\boldsymbol{a}_{n}$ at time slot $t$ as follows: 
	\begin{equation} 
	\label{eq:Convergence_action_probability}
	\begin{split}
	P(\boldsymbol{a}_{n}) = \theta_{n}^{\boldsymbol{a}_{n}} (1-\theta_{n})^{1-\boldsymbol{a}_{n}} \;\;\;\;\;\;\;\;\;\;\;\;\;\;\;\;\;\;\;\;\;\;\;\\
	\;\;\;\;\;\;\;\;\;\;\;\;\;\;\;	= {\boldsymbol{a}_{n}} \log \theta_{n} + ({1-\boldsymbol{a}_{n}}) \log (1-\theta_{n}).
	\end{split}
	\end{equation} 
	Now, for a single sample the policy gradient estimator is defined as follows:
	\begin{equation} 
	\label{eq:Convergence_policy_gradient_estimator}
	\begin{split}
	\frac{\hat{\partial}}{\partial \theta_{n}} J(\theta_{n}) = r_{n}(\boldsymbol{a}_{1}, \dots, \boldsymbol{a}_{N}) \frac{\partial}{\partial \theta_{n}} \log P(\boldsymbol{a}{1}, \dots, \boldsymbol{a}_{N}) \;\;\;\;\;\;\;\;\;\;\;\;\;\\
	= r_{n}(\boldsymbol{a}_{1}, \dots, \boldsymbol{a}_{N}) \frac{\partial}{\partial \theta_{n}} \sum_{\forall n \in \mathcal{N}} {\boldsymbol{a}_{n}} \log \theta_{n} + ({1-\boldsymbol{a}_{n}}) \log (1-\theta_{n}) \\
	= r_{n}(\boldsymbol{a}_{1}, \dots, \boldsymbol{a}_{N}) \frac{\partial}{\partial \theta_{n}} ( {\boldsymbol{a}_{n}} \log \theta_{n} + ({1-\boldsymbol{a}_{n}}) \log (1-\theta_{n})) \;\;\;\;\;\;\\
	= r_{n}(\boldsymbol{a}_{1}, \dots, \boldsymbol{a}_{N}) (\frac{\boldsymbol{a}_{n}}{\theta_{n}} - \frac{(1-\boldsymbol{a}_{n})}{(1-\theta_{n})}) \;\;\;\;\;\;\;\;\;\;\;\;\;\;\;\;\;\;\;\;\;\;\;\;\;\;\;\;\;\;\;\;\;\;\;\;\;\\
	= r_{n}(\boldsymbol{a}_{1}, \dots, \boldsymbol{a}_{N}) (2\boldsymbol{a}_{n}-1),\;
	\text{for $\theta_{n} = \frac{1}{2}$}. \;\;\;\;\;\;\;\;\;\;\;\;\;\;\;\;\;\;\;\;\;\;\;\;\;\;\;\;
	\end{split}
	\end{equation}
	Now, the expected reward for the $N$ agents is defined by $\mathbb{E}[r_{n}] = \sum_{\forall n \in \mathcal{N}} r_{n}(\boldsymbol{a}_{1}, \dots, \boldsymbol{a}_{N}) (\frac{1}{2})^N$, and using $r_{n}(\boldsymbol{a}_{1}, \dots, \boldsymbol{a}_{N}) = 1 |(\boldsymbol{a}_{1}= \dots = \boldsymbol{a}_{N})$, we get $\mathbb{E}[r_{n}] = (\frac{1}{2})^N$. Therefore, the expectation of the gradient estimation is $ \mathbb{E}[\frac{\hat{\partial}}{\partial \theta_{n}} J(\theta_{n})] = \frac{\partial}{\partial \theta_{n}}J(\theta_{n}) = (\frac{1}{2})^N$. The variance of the estimated gradient can be calculated as follows:
	\begin{equation} 
	\label{eq:Convergence_variance}
	\begin{split}
	\mathbb{V}\big[\frac{\hat{\partial}}{\partial \theta_{n}} J(\theta_{n})\big] = \mathbb{E}\big[\frac{\hat{\partial}}{\partial \theta_{n}} J^2(\theta_{n})\big] - \left(\mathbb{E}\big[\frac{\hat{\partial}}{\partial \theta_{n}} J(\theta_{n})\big]\right)^2 \\ = \left(\frac{1}{2}\right)^N - \left(\frac{1}{2}\right)^{2N}. \;\;\;\;\;\;\;\;\;\;\;\;\;\;\;\;\;\;\;\;\;\;\;\;\;\;
	\end{split}
	\end{equation} 
	From \eqref{eq:Convergence_relation_policy_grdient}, we can analyze $P((\hat{\nabla}_{\theta_{n}} J(\theta_{n}), \nabla_{\theta_{n}} J(\theta_{n})) > 0)$ and get
	\begin{equation}
	\label{eq:Convergence_final}
	\begin{split}
	P\left(\hat{\nabla}_{\theta_{n}} J(\theta_{n}), \nabla_{\theta_{n}} J(\theta_{n})\right) = \left(\frac{1}{2}\right)^N \sum_{\forall n \in \mathcal{N}} \frac{\hat{\partial}}{\partial \theta_{n}} J(\theta_{n}).
	\end{split}
	\end{equation} 
	Therefore, $P((\hat{\nabla}_{\theta_{n}} J(\theta_{n}), \nabla_{\theta_{n}} J(\theta_{n})) > 0) = (\frac{1}{2})^N$, which implies that the gradient step moves in the correct direction and decreases exponentially with an increasing number of agents.	
\end{proof}

\ifCLASSOPTIONcaptionsoff
  \newpage
\fi

\begin{IEEEbiography}[{\includegraphics[width=1in,height=1.25in,clip,keepaspectratio]{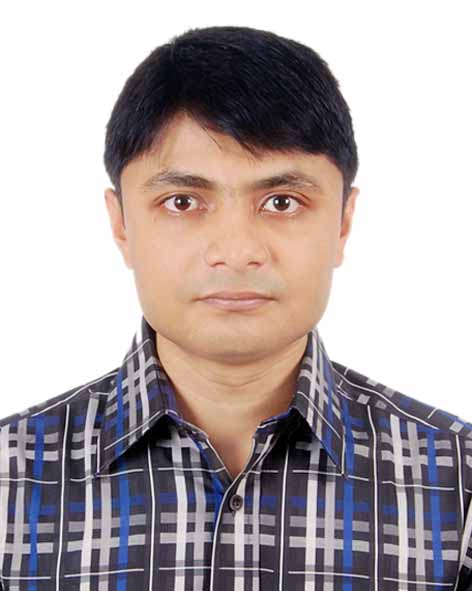}}]{Md.~Shirajum~Munir}
	(Graduate Student Member, IEEE) received the B.S. degree in computer science and engineering from Khulna
	University, Khulna, Bangladesh, in 2010. He is currently pursuing the Ph.D. degree in computer science and engineering at Kyung Hee University, Seoul,
	South Korea. He served as a Lead Engineer with the Solution	Laboratory, Samsung Research and Development Institute, Dhaka, Bangladesh, from 2010 to 2016. His current research interests include IoT network management, mobile edge computing, software-defined networking, smart grid, and machine learning.
\end{IEEEbiography}
\begin{IEEEbiography}[{\includegraphics[width=1in,height=1.25in,clip,keepaspectratio]{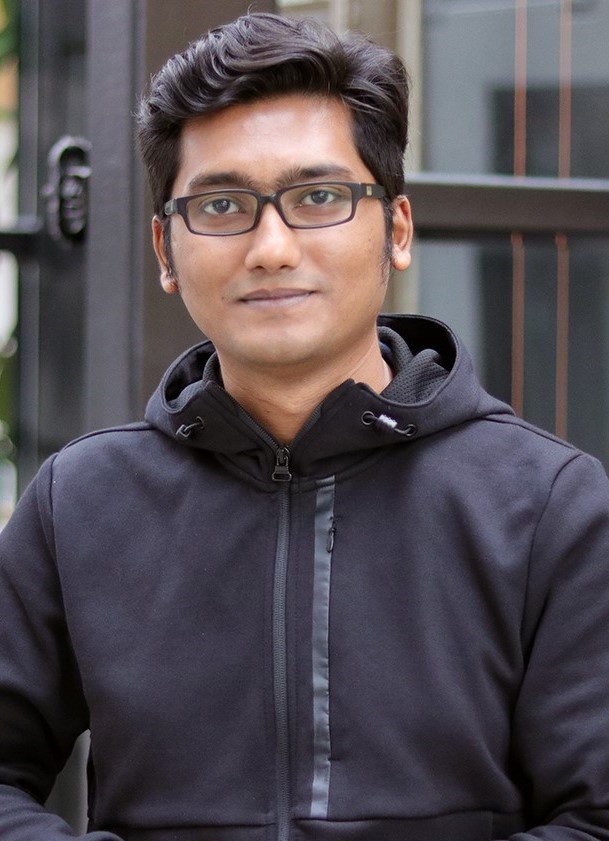}}]{Sarder~Fakhrul~Abedin} (Member, IEEE) received his B.S. degree in Computer Science from Kristianstad University, Kristianstad, Sweden, in 2013. He received his Ph.D. degree in Computer Engineering from Kyung Hee University (KHU), South Korea in 2020. He served as Postdoctoral Researcher at the Department of Computer Science and Engineering, KHU, and also at Mid Sweden University (MIUN), Sweden. Currently, he is an Assistant Professor with the Department of Information Systems and Technology, MIUN.  His research interests include Internet of Things (IoT) network management, Fog computing, Machine learning, Industrial 5G and Wireless networking.
\end{IEEEbiography}
\begin{IEEEbiography}[{\includegraphics[width=1in,height=1.25in,clip,keepaspectratio]{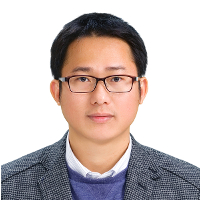}}]{Nguyen~H.~Tran}
	(S'10-M'11-SM'18)  received the	B.S. degree from the Ho Chi Minh City University
	of Technology, Ho Chi Minh City, Vietnam, in 2005, and the Ph.D. degree in electrical and computer engineering from Kyung Hee University, Seoul, South Korea, in 2011. Since 2018, he has been with the School of Computer Science, University of Sydney, Sydney,	NSW, Australia, where he is currently a Senior Lecturer. He was an Assistant Professor with the Department of Computer Science and Engineering, Kyung Hee University, from 2012 to 2017. His current research interests include applying analytic techniques of optimization, game theory, and stochastic modeling to cutting-edge applications, such as cloud and mobile edge computing, data centers, heterogeneous wireless networks, and big data for networks. Dr. Tran was a recipient of the Best KHU Thesis Award in Engineering	in 2011 and the Best Paper Award of IEEE ICC 2016. He has been an Editor of the IEEE TRANSACTIONS ON GREEN COMMUNICATIONS AND	NETWORKING since 2016 and served as the Editor of the 2017 Newsletter of Technical Committee on Cognitive Networks on Internet of Things.
\end{IEEEbiography}
\begin{IEEEbiography}[{\includegraphics[width=1in,height=1.25in,clip,keepaspectratio]{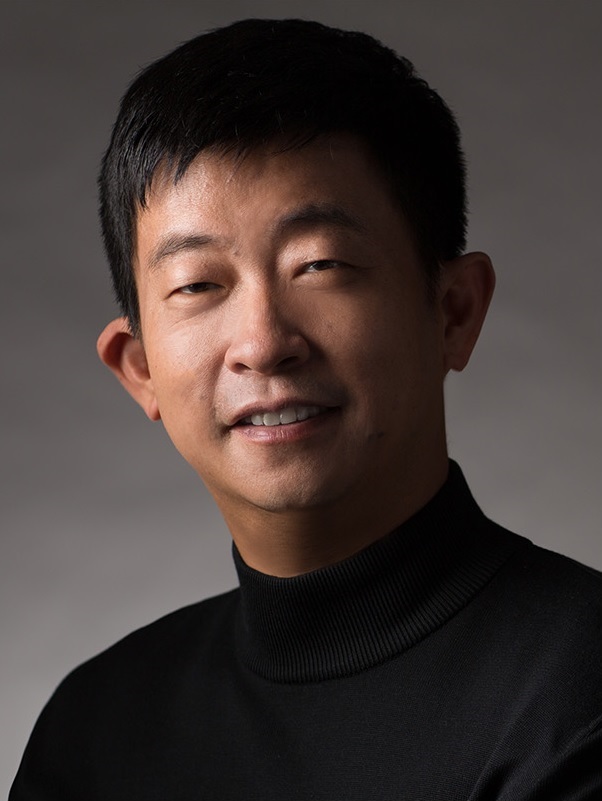}}]{Zhu~Han}
	(S'01-M'04-SM'09-F'14) received the B.S. degree in electronic engineering from Tsinghua University, in 1997, and the M.S. and Ph.D. degrees in electrical and computer engineering from the University of Maryland, College Park, in 1999 and 2003, respectively. 
	
	From 2000 to 2002, he was an R\&D Engineer of JDSU, Germantown, Maryland. From 2003 to 2006, he was a Research Associate at the University of Maryland. From 2006 to 2008, he was an assistant professor at Boise State University, Idaho. Currently, he is a John and Rebecca Moores Professor in the Electrical and Computer Engineering Department as well as in the Computer Science Department at the University of Houston, Texas. His research interests include wireless resource allocation and management, wireless communications and networking, game theory, big data analysis, security, and smart grid.  Dr. Han received an NSF Career Award in 2010, the Fred W. Ellersick Prize of the IEEE Communication Society in 2011, the EURASIP Best Paper Award for the Journal on Advances in Signal Processing in 2015, IEEE Leonard G. Abraham Prize in the field of Communications Systems (best paper award in IEEE JSAC) in 2016, and several best paper awards in IEEE conferences. Dr. Han was an IEEE Communications Society Distinguished Lecturer from 2015-2018, AAAS fellow since 2019 and ACM distinguished Member since 2019. Dr. Han is $1\%$ highly cited researcher since 2017 according to Web of Science. Dr. Han is also the winner of 2021 IEEE Kiyo Tomiyasu Award, for outstanding early to mid-career contributions to technologies holding the promise of innovative applications, with the following citation: ``for contributions to game theory and distributed management of autonomous communication networks."
	
\end{IEEEbiography}
\begin{IEEEbiography}[{\includegraphics[width=1in,height=1.25in,clip,keepaspectratio]{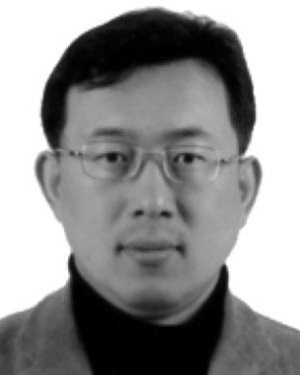}}]{Eui-Nam Huh} (M'13)
	received the B.S. degree from Busan National University, South Korea, the master's degree in computer science from The	University of Texas, USA, in 1995, and the Ph.D.
	degree from the Ohio University, USA, in 2002. He is currently with Kyung Hee University,	South Korea, as a Professor with the Department of Computer Science and Engineering. His research interests include cloud computing, the Internet of
	Things, the future Internet, distributed real time systems, mobile computing, big data, and security. He is also Review Board	of the National Research Foundation of Korea. He has also served many community services for ICCSA, WPDRTS/IPDPS, APAN Sensor Network Group, ICUIMC, ICONI, APIC-IST, ICUFN, and SoICT as various types of
	chairs. He is a vice-chairman of Cloud/Bigdata Special Technical Group of TTA and an Editor of ITU-T SG13 Q17.
\end{IEEEbiography}
\begin{IEEEbiography}[{\includegraphics[width=1in,height=1.25in,clip,keepaspectratio]{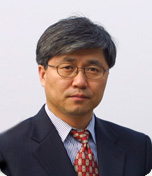}}]{Choong~Seon~Hong}
	(S'95-M'97-SM'11)
	 received the B.S. and M.S. degrees in electronic engineering from Kyung Hee University, Seoul, South Korea, in 1983 and 1985, respectively, and the Ph.D. degree from Keio University, Japan, in 1997. In 1988, he joined KT, where he was involved in broadband networks as a Member of Technical Staff. Since 1993, he has been with Keio University. He was with the Telecommunications Network Laboratory, KT, as a Senior Member of Technical Staff and as the Director of the Networking Research Team until 1999. Since 1999, he has been a Professor with the Department of Computer Science and Engineering, Kyung Hee University. His research interests include future Internet, ad hoc networks, network management, and network security. He is a member of the ACM, the IEICE, the IPSJ, the KIISE, the KICS, the KIPS, and the OSIA. Dr. Hong has served as the General Chair, the TPC Chair/Member, or an Organizing Committee Member of international conferences such as NOMS, IM, APNOMS, E2EMON, CCNC, ADSN, ICPP, DIM, WISA, BcN, TINA, SAINT, and ICOIN. He was an Associate Editor of the IEEE TRANSACTIONS ON NETWORK AND SERVICE MANAGEMENT, and the IEEE JOURNAL OF COMMUNICATIONS AND NETWORKS. He currently serves as an Associate Editor of the International Journal of Network Management, and an Associate Technical Editor of the IEEE Communications Magazine.
\end{IEEEbiography}

\end{document}